\documentclass[10pt,twocolumn,twoside]{IEEEtran}
\usepackage{amsfonts,dsfont,amssymb,amsbsy,amsmath,paralist,theorem,bm,ifthen,color}
\usepackage[pdfstartview=FitH,bookmarksnumbered,unicode,bookmarksopen=true]{hyperref}
\usepackage{graphicx}
\usepackage{epstopdf}
\usepackage{pifont}
\usepackage{relsize}
\usepackage{mathrsfs}
\usepackage{tcolorbox}
\usepackage{bm}
\usepackage{stfloats}
\usepackage{url}
\usepackage[noadjust]{cite}
\usepackage{cases,booktabs}
\usepackage{graphicx,algorithmic,algorithm,relsize}
\usepackage[justification=centering]{caption} 

\usepackage[table]{xcolor}
\usepackage{array,tabularx, multirow, multicol, booktabs, hyperref}
\usepackage{threeparttable}
\usepackage{tablefootnote}

\newcommand\pin{\ensuremath{{\rm Pin}}}

\newcommand\wb{\ensuremath{{\bf w}}}
\newcommand\yb{\ensuremath{{\bf y}}}

\newcommand\hb{\ensuremath{{\bf h}}}

\newcommand\gb{\ensuremath{{\bf g}}}
\newcommand\Ib{\ensuremath{{\bf I}}}

\newcommand\Xb{\ensuremath{{\bf X}}}
\newcommand\xb{\ensuremath{{\bf x}}}

\newcommand\nb{\ensuremath{{\bf n}}}

\newcommand\tb{\ensuremath{{\bm t}}}

\newcommand\vb{\ensuremath{{\bf v}}}

\newcommand\psib{\ensuremath{{\bm \psi}}}

\def\CN{{{\mathcal{CN}}}}

\newcommand\E{\ensuremath{{\mathbb{E}}}}

\newcommand\Cs{\ensuremath{{\mathbb{C}}}}

\newcommand\maxn  {\ensuremath{{\rm maximize}}}
\newcommand\st    {\ensuremath{{\rm subject~to}}}

\newcommand\Nset  {\ensuremath{{\mathcal{N}}}}
\newcommand\Mset  {\ensuremath{{\mathcal{M}}}}

\newcommand\Hf{\ensuremath{{\mathsf{H}}}}

\graphicspath{{fig/}}

\definecolor{sectbg}{RGB}{235,240,255} 
\definecolor{sectbg2}{RGB}{236, 236, 236} 
\definecolor{green}{RGB}{34	195	46}
\definecolor{red}{RGB}{220 0 0}

\usepackage{graphicx} 

\allowdisplaybreaks[4]

\title{Generalized Pinching-Antenna Systems: A Tutorial on Principles, Design Strategies, and Future Directions}

\author{Yanqing Xu, Jingjing Cui, Yongxu Zhu, Zhiguo Ding, Tsung-Hui Chang, Robert Schober, Vincent W.S. Wong, Octavia A. Dobre, George K. Karagiannidis, H. Vincent Poor, and Xiaohu You
\\ 
     \thanks{\smaller[1] Y. Xu is with the School of Science and Engineering, The Chinese University of Hong Kong, Shenzhen, 518172, China (email: xuyanqing@cuhk.edu.cn).}
     \thanks{\smaller[1] J. Cui is with the School of Information Science and Technology, Southwest Jiaotong University, Chengdu 610031, China (e-mail: jingjing.cui@swjtu.edu.cn).}
     \thanks{\smaller[1] Y. Zhu and X. You are with the National Mobile Communications Research Laboratory, Southeast University, Nanjing 210096, China. (e-mail: yongxu.zhu@seu.edu.cn; xhyu@seu.edu.cn).}
     \thanks{\smaller[1] Z. Ding is with the School of Electrical \& Electronic Engineering, Nanyang Technological University, 639798, Singapore  (e-mail: zhiguo.ding@ntu.edu.sg).}
     \thanks{\smaller[1] T.-H. Chang is with the School of Artificial Intelligence, The Chinese University of Hong Kong, Shenzhen, 518172, China (email: changtsunghui@cuhk.edu.cn).}      
     \thanks{\smaller[1] R. Schober is with the Institute for Digital Communications, Friedrich-Alexander-University Erlangen-N\"{u}rnberg (FAU), 91054 Erlangen, Germany (e-mail: robert.schober@fau.de).}
      \thanks{\smaller[1] V. W.S. Wong is with the Department of Electrical and Computer Engineering, The University of British Columbia, Vancouver, BC, V6T 1Z4, Canada    (e-mail: vincentw@ece.ubc.ca).}
     \thanks{\smaller[1]  O. A. Dobre is with the Faculty of Engineering and Applied Science, Memorial University, St. Johns, NL A1C 5S7, Canada (e-mail: odobre@mun.ca).}
     \thanks{\smaller[1] G. K. Karagiannidis is with the Department of Electrical and Computer Engineering, Aristotle University of Thessaloniki, 541 24 Thessaloniki, Greece (e-mail: geokarag@auth.gr).}
     \thanks{\smaller[1] H. V. Poor is with the Department of Electrical and Computer Engineering, Princeton University, Princeton, NJ 08544 USA (e-mail: poor@princeton.edu).}

}

\date{\today}

\begin{document}

\maketitle

\begin{abstract}
Pinching-antenna systems have emerged as a novel and transformative flexible-antenna architecture for next-generation wireless networks. They offer unprecedented flexibility and spatial reconfigurability by enabling dynamic positioning and activation of radiating elements along a signal-guiding medium (e.g., dielectric waveguide), which is not possible with conventional fixed antenna systems.
In this paper, we introduce the concept of generalized pinching antenna systems, which retain the core principle of creating localized radiation points on demand, but can be physically realized with a variety of technologies. These include implementations based on dielectric waveguides, leaky coaxial cables, surface-wave guiding structures, and other types of media, employing different feeding methods and activation mechanisms (e.g., mechanical, electronic, or hybrid). Despite differences in their physical realizations, they all share the same inherent ability to form, reposition, or deactivate radiation sites as needed, enabling user-centric and dynamic coverage.
We first describe the underlying physical mechanisms of representative generalized pinching-antenna realizations and their associated wireless channel models, highlighting their unique propagation and reconfigurability characteristics compared with conventional antennas. 
Then, using dielectric waveguide-based pinching antennas as primary examples, we review several representative pinching-antenna system architectures, ranging from single- to multiple-waveguide configurations, and discuss advanced design strategies tailored to these flexible deployments. Furthermore, we examine their integration with emerging wireless technologies to enable synergistic, user-centric solutions. Finally, we identify key open research challenges and outline future directions, charting a pathway toward the deployment of generalized pinching antennas in next-generation wireless networks.

\end{abstract}


\section{Introduction}

\subsection{Background and Motivation}

The evolution of wireless networks is entering a new era with the advent of sixth-generation (6G) wireless systems. Exceeding the capabilities of previous generations, 6G is expected to support various new applications and use cases, including immersive extended reality, holographic telepresence, intelligent robotics, edge artificial intelligence (AI), generative AI, and large-scale Internet of Things (IoT) deployments \cite{you2021towards,zhang20196g,ITU-R_2023,zhang2024generative1,zhang2024interactive2,zhang2024generative3}. These new use cases have different requirements on data rates, latency, reliability, and connectivity density. Meeting these diverse requirements requires innovations across the protocol stack. In the physical layer, one of the most critical components is the multi-antenna subsystem, which directly influences system throughput, spatial coverage, and energy efficiency \cite{xu2025distributed,larsson2014massive}. The wireless industry has traditionally relied on fixed or semi-static multi-antenna systems, such as conventional multiple-input multiple-output (MIMO) arrays. These technologies exploit spatial diversity, enhance spectral efficiency, and combat multipath fading, thereby enabling significant performance improvements in fourth-generation (4G) and fifth-generation (5G) wireless networks.
However, a major limitation of traditional multi-antenna solutions is their limited reconfigurability. Once the antennas are fabricated and installed, their physical and electromagnetic properties are largely fixed. Consequently, these systems lack the flexibility to adapt to changes in propagation conditions, spectrum usage, and user locations. Looking ahead, future wireless networks will become increasingly dense and dynamic. They will need to cope with users' mobility, reconfigurable environments, and highly variable spectrum allocations. The limited reconfigurability of traditional antennas may lead to significant performance bottlenecks in such scenarios.

These challenges have catalyzed a major shift toward flexible antenna architectures.
Representative examples include reconfigurable intelligent surfaces (RISs), also known as intelligent reconfigurable surfaces (IRSs) \cite{tang2020wireless,wu2019towards}, fluid antennas \cite{wong2020fluid,new2024tutorial}, and movable antenna systems \cite{ma2023mimo,zhang2024channel}. These systems have attracted significant research attention due to their ability to dynamically modify wireless channels and enhance communication performance. RISs electronically adjust the reflection coefficients of large arrays of passive elements to enable programmable propagation environments. Fluid antenna systems use the physical properties of conducting fluids to alter the antennas' positions or geometry, enabling agile adaptation to changing signal conditions. Movable antenna systems provide flexibility by mechanically displacing antenna elements within a predefined region to support position-dependent channel optimization.
While these techniques bring significant advances, they cannot directly address the challenging issue of line-of-sight (LoS) blockage. Each of the aforementioned techniques has its own practical limitations. RIS-based systems may suffer from double attenuation in the transmitter-to-RIS and RIS-to-receiver paths. This can result in substantial path loss, especially under non-line-of-sight (NLoS) conditions. Both fluid and movable antennas generally restrict their movement to a few wavelengths, limiting their ability to create new LoS channels and mitigate large-scale path loss when user locations vary significantly. These limitations may affect their performance in highly dynamic or densely deployed network scenarios.

To address the aforementioned challenges, the research community is exploring advanced  reconfigurable antenna technologies beyond current paradigms. Among the emerging technologies, pinching-antenna systems have recently received significant attention as a promising new approach. They have the potential to enhance flexibility, reduce implementation cost, provide fine-grained control, and improve integration into next-generation wireless networks.

\begin{figure*}
    \centering
    \includegraphics[width=0.98\linewidth]{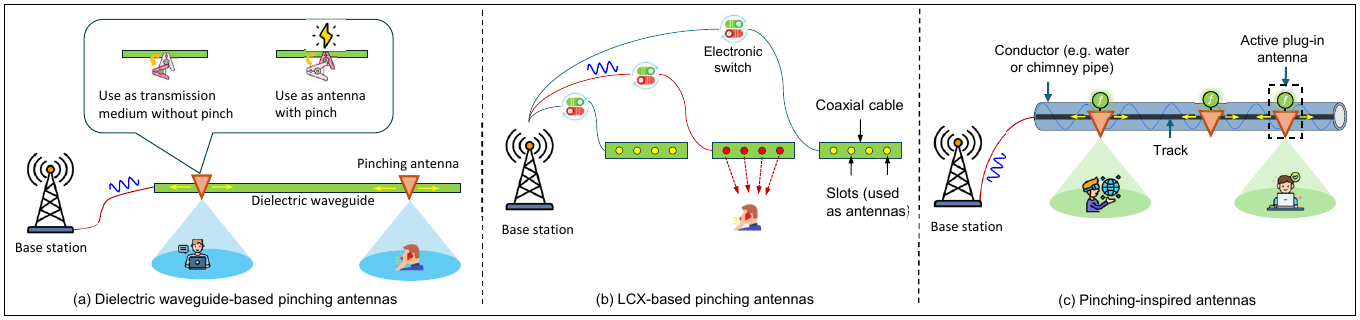}
        \captionsetup{justification=justified, singlelinecheck=false, font=small}	
        \caption{Conceptual illustration of generalized pinching-antenna systems: (a) Dielectric waveguide-based pinching antennas use discrete dielectric particles or clips to enable localized radiation along a dielectric waveguide; (b) LCX-based pinching antennas employ periodic slots on coaxial cables, with electronic switches enabling on-demand signal leakage through selective segment activation; and (c) Pinching-inspired antennas, in which active antennas are strategically placed along a transmission surface to support user-centric and on-demand wireless coverage.}
        \label{fig: gpa}
\end{figure*}

\subsection{Generalized Pinching-Antenna Systems: Basic Principles and Key Advantages}
Generalized pinching-antenna systems represent a class of antenna architectures in which radio frequency (RF) signals are first guided along a physical medium and then are selectively radiated into free space at configurable radiation points, creating wireless links near the users. The fundamental principle is to enable localized electromagnetic energy leakage at specific locations, which allows wireless coverage to be tailored according to user demand and scenario requirements. This approach overcomes the limitations of traditional fixed antenna installations by offering better adaptability and user-centric deployment in wireless networks. In Fig. \ref{fig: gpa}, three representative examples of generalized pinching-antenna systems are presented.
This includes NTT DOCOMO's dielectric waveguide-based pinching-antenna system \cite{suzuki2022pinching}, where the radiating elements are physically pinched (i.e., attached) to a dielectric waveguide using discrete dielectric particles or clips (see Fig. \ref{fig: gpa}(a)). Each pinch disrupts signals, which are propagated along the waveguide, causing controlled energy leakage at the desired position, effectively ``creating'' an antenna. This process is reversible and reconfigurable. That is, the radiating elements can be deployed, repositioned, or removed flexibly, enabling highly adaptive coverage with minimal infrastructure modification \cite{ding2025flexible}.

Leaky coaxial cables (LCXs) are another important example of generalized pinching antennas. Conventional LCX antennas radiate electromagnetic signals through periodic slots in coaxial cables, providing reliable wireless coverage in tunnels, subway systems, mines, and other confined or elongated environments \cite{torrance1996multi,wang2001theory,wang2016modeling,wu2016performance}.
Compared to DOCOMO's pinching antennas, conventional LCX antennas lack flexibility and user-centric control, since their radiation points are fixed and cannot be dynamically reconfigured to satisfy the  changing coverage requirement. However, by dividing an LCX cable into multiple segments and allowing these segments to be selectively activated or deactivated, e.g., through electronic switching, radiating sections can be dynamically tailored to user locations or coverage requirements, which yields an LCX-based pinching-antenna system as shown in Fig. \ref{fig: gpa}(b). This localized and user-centric radiation mechanism parallels the on-demand flexibility of dielectric waveguide-based pinching antennas, albeit through a different physical implementation.

The key idea of pinching antennas can also be used to design various pinching-inspired antenna arrays. Another representative example is shown in Fig. \ref{fig: gpa}(c), where the signals are first guided through a metal pipe via the surface-wave mode and then radiated in an on-demand manner by applying an active antenna close to the users \cite{polo2013electromagnetic,gao2015spoof}. 
Such systems can operate similarly to power track systems, where each pinching antenna functions like a plug-in module that can be flexibly deployed along the medium.
Unlike surface-wave communication, which is non-radiating, the use of such pinching-inspired antenna arrays ensures that electromagnetic signals are radiated into free space at the chosen locations. Another example draws from the concept of radio stripes \cite{shaik2021mmse,mishra2021millimeter}, in which a series of radiating sites are embedded along the extended conductive structures. In this approach, signals are transmitted with low attenuation through the radio stripe to positions that are close to the intended users, where they are subsequently radiated into free space. Integrating active antennas, directional couplers, or segmented elements enables these antennas to activate radiation on demand and adapt dynamically to coverage and user demands.

Collectively, generalized pinching-antenna systems provide a flexible deployment paradigm in which radiating elements can be dynamically activated, repositioned, or deactivated along a transmission medium. This capability allows wireless coverage to adapt to user locations and changing environmental conditions without requiring extensive physical infrastructure or fixed antenna layouts. As illustrated in Fig. \ref{fig: gpa advantages}, such systems depart from conventional antenna architectures by supporting on-demand formation of radiation sites, thereby facilitating user-centric and scalable communication. This distinctive reconfigurability introduces several key technical and practical advantages for next-generation wireless networks:

\begin{figure*}
    \centering
    \includegraphics[width=0.98\linewidth]{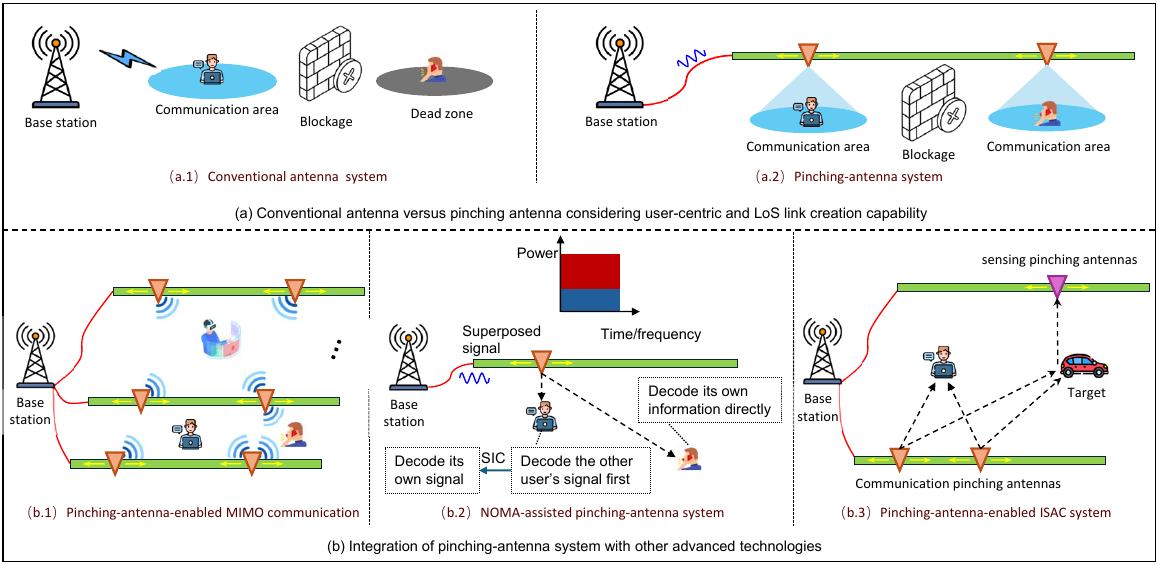}
        \captionsetup{justification=justified, singlelinecheck=false, font=small}	
        \caption{Illustration of the advantages of generalized pinching-antenna systems using the dielectric waveguide-based implementation as an example: (a) Comparison between conventional antenna systems and pinching-antenna systems in terms of user-centric and LoS link creation; and (b) Integration of pinching-antenna system with other advanced wireless technologies.}
        \label{fig: gpa advantages}
\end{figure*}

\begin{itemize}
    \item {\bf Highly Flexible and Scalable Configuration:} Generalized pinching-antenna systems offer exceptional reconfigurability by allowing antennas to be dynamically added, repositioned, or removed with minimal mechanical effort. This flexibility facilitates real-time adaptation to user locations, mobility patterns, and coverage requirements. Unlike conventional arrays with fixed geometries, pinching antennas support fully customizable topologies, where the array size, spatial distribution, and angular beam direction can all be tailored and reconfigured after deployment.
    \item {\bf User-Centric and Robust LoS Connectivity:} Generalized pinching-antenna systems enable flexible deployment and dynamic repositioning of radiating elements along the transmission medium, making it possible to establish strong LoS links tailored to user locations and adapt to mobility in real time. This user-centric capability ensures enhanced coverage and reliable performance, particularly in high-frequency (e.g., millimeter wave (mmWave) and terahertz (THz)) bands and challenging environments such as dense urban areas or obstructed indoor scenarios, making the system ideal for dynamic and user-driven wireless service provisioning.
    \item {\bf Support for Near-Field Communication:} The ability to deploy/activate pinching antennas over extended physical apertures facilitates operation in the near-field region, where spherical wavefronts dominate signal propagation. This allows for distance-dependent beamforming, enhanced spatial resolution, and highly focused energy transmission, enabling advanced functionalities not achievable with traditional far-field antenna systems.
    \item {\bf Integration with Advanced Wireless Technologies:} Generalized pinching antennas can be seamlessly combined with emerging wireless technologies such as MIMO, non-orthogonal multiple access (NOMA), hybrid beamforming, integrated sensing and communications (ISAC), wireless power transfer, and edge AI. This integration enables new transmission architectures and helps address the diverse requirements of next-generation wireless communications.
    \item {\bf Practicality, Cost Effectiveness, and Simplicity:} Generalized pinching-antenna systems offer a highly practical and economically viable solution by leveraging low-cost transmission media and readily manufacturable pinching units. Unlike conventional antenna arrays that require complex RF chains for each radiating element, pinching antennas significantly reduce hardware complexity by enabling signal radiation directly from waveguide pinches. This minimalist architecture not only reduces material and deployment costs but also simplifies system integration, maintenance, and scalability, making it particularly attractive for large-scale, user-centric, and flexible wireless deployments in both indoor and outdoor scenarios.
\end{itemize}

\begin{table*}[!t]
\centering
\caption{Qualitative comparison of pinching-antenna systems, RIS, and fluid/movable antennas.}
\label{tab:comparison_pinching_ris_fluid}
\renewcommand{\arraystretch}{2.1}
\begin{tabular}{
  >{\centering\arraybackslash}p{2.9cm}
  >{\arraybackslash}p{4.8cm}
  >{\arraybackslash}p{3.7cm}
  >{\arraybackslash}p{4.9cm}}
\hline\hline
\textbf{Aspect} &
\multicolumn{1}{c}{\textbf{Generalized pinching antennas}} &
\multicolumn{1}{c}{\textbf{RIS}} &
\multicolumn{1}{c}{\textbf{Fluid/movable antennas}} \\
\hline
\multirow{1.6}{*}{\textbf{Signal propagation}} 
& Wired segment $+$ short-distance wireless link 
& Two-hop BS–RIS–user wireless link
& Single-hop wireless link with reconfigurable radiating point \\

\multirow{2.2}{*}{\textbf{Path-loss behaviour}}
& Wired segment exhibits low loss along guided structures, while short wireless link incurs only small free-space path loss
& Experiences two-hop path loss; performance sensitive to BS–RIS–user topology
& Similar to conventional antennas, path loss is mainly determined by user–BS link distance, with gains from local fading diversity \\

\multirow{1.6}{*}{\textbf{Reconfiguration scale}}
& Large-scale along guiding media (tens of metres, many radiating points) 
& Panel-scale on a fixed surface (element-wise reflection control)
& Small-scale at wavelength-level antenna reconfiguration and movement\\

\multirow{2.2}{*}{\textbf{Hardware / deployment}}
& Dielectric waveguides, LCX, or similar media; few RF chains feeding many radiating points 
& Large passive panel on existing surfaces; no extra RF chains but many tunable elements 
& Fluid antennas use tunable conductive liquids in small cavities, while movable antennas rely on compact mechanical actuators.\\
\hline\hline
\end{tabular}
\end{table*}

\begin{table*}[!t]
\centering
\caption{Comparison between this work and representative references on pinching-antenna systems.}
\label{tab:comparison_previous_works}
\renewcommand{\arraystretch}{1.6}
\begin{tabular}{
  |c|c|
  >{\centering\arraybackslash}p{1.2cm}|
  >{\centering\arraybackslash}p{1.3cm}|
  >{\centering\arraybackslash}p{1.3cm}|
  >{\centering\arraybackslash}p{2.0cm}|
  >{\centering\arraybackslash}p{2.6cm}|
  >{\centering\arraybackslash}p{3.0cm}|}
\hline
\multirow{4}{*}{Reference} & \multirow{4}{*}{Paper Type} &
\multicolumn{3}{c|}{Pinching-antenna realizations} &
\multicolumn{3}{c|}{Main elements covered} \\
\cline{3-8}
& &
\begin{tabular}[c]{@{}c@{}}Dielectric\\ waveguide-\\based\end{tabular} &
\begin{tabular}[c]{@{}c@{}}LCX-\\based\end{tabular} &
\begin{tabular}[c]{@{}c@{}}Pinching-\\inspired\end{tabular} &
\begin{tabular}[c]{@{}c@{}}Channel modelling\\ and system\\ optimization\end{tabular} &
\begin{tabular}[c]{@{}c@{}}Impact of key system\\ parameters (e.g., \\  attenuation, blockage)\end{tabular} &
\begin{tabular}[c]{@{}c@{}}Integration with advanced\\ techniques, applications,\\ and open directions\end{tabular} \\
\hline
\cite{ding2025flexible}         & Technical & \ding{51} & \ding{55} & \ding{55} & \ding{51} & \ding{55} & \ding{51} \\
\cite{yang2025pinching}        & Magazine  & \ding{51} & \ding{55} & \ding{55} & \ding{55} & \ding{55} & \ding{51} \\
\cite{liu2025pinching}         & Magazine  & \ding{51} & \ding{55} & \ding{55} & \ding{55} & \ding{55} & \ding{51} \\
\cite{liu2025pinching-tutorial}& Tutorial  & \ding{51} & \ding{55} & \ding{55} & \ding{51} & \ding{55} & \ding{51} \\
This work                      & Tutorial  & \ding{51} & \ding{51} & \ding{51} & \ding{51} & \ding{51} & \ding{51} \\
\hline
\end{tabular}
\end{table*}

An illustration of the advantages of generalized pinching antennas is provided in Fig.~\ref{fig: gpa advantages}. As shown from the figure, generalized pinching-antenna systems operate in a different regime compared to both RISs and fluid/movable antennas. RISs reshape the small-scale wireless channel using a fixed reflecting surface by tuning a large number of passive elements, while fluid/movable antennas reconfigure a radiating point within a small region to exploit local fading diversity. In both cases, the large-scale propagation geometry and the base station (BS)–user path loss remain essentially unchanged. In contrast, generalized pinching antennas exploit low-loss guiding media (dielectric waveguides, LCX, or surface-wave conductors) to realize large-scale channel reconfiguration employing a small number of feeder points, creating, removing, or repositioning radiating spots over tens of meters along installed structures, while incurring only modest RF-chain and cabling overhead. A qualitative comparison between these flexible-antenna techniques is summarized in Table \ref{tab:comparison_pinching_ris_fluid}.

\begin{table*}[t]
\centering
\caption{Outline of the paper.}
\renewcommand{\arraystretch}{1.6}
\setlength{\tabcolsep}{6pt}
\begin{tabular}{p{0.48\linewidth} p{0.48\linewidth}}
\hline \hline
\multicolumn{2}{c}{\textbf{Section I. Introduction}}\\
A. Motivation and Background & B. Generalized Pinching-Antenna Systems \\
C. Contributions and Organization of the Paper &   \\
\multicolumn{2}{c}{\textbf{Section II. Fundamentals of Generalized Pinching Antennas}}\\
A. Dielectric Waveguide-Based Pinching Antennas & B. LCX-Based Pinching Antennas\\
C. Pinching-Inspired Antenna Systems &  \\
\multicolumn{2}{c}{\textbf{Section III. Representative System Architectures and Design Strategies for Pinching-Antenna Systems}}\\
A. Single-Waveguide Systems (Single/Multiple Antennas per Waveguide) & B. Multi-Waveguide Systems (Single/Multiple Antennas per Waveguide) \\
C. Pinching-Antenna System Design under Random LoS and NLoS Channels & D. Uplink Pinching-Antenna Systems \\
E. Control and Implementation of Pinching-Antenna Positioning &  \\
\multicolumn{2}{c}{\textbf{Section IV. Broader Horizons: Versatile and Emerging Applications of Pinching Antennas}}\\
A. Enabling ISAC with Pinching Antennas & B. Facilitating Cooperative Communications with Pinching Antennas \\
C. Synergizing Pinching Antennas with Other Advanced Techniques &  \\
\multicolumn{2}{c}{\textbf{Section V. Charting the Future: Open Questions and Research Directions}}\\
A. Pinching-Antenna System Design under LoS Blockage & B. Realization and Optimization of Efficient Pinching-Antenna Radiators \\
C. Pinching-Antenna-Enabled Near-Field Communications & D. Accurate and Realistic Electromagnetic Modeling of Pinching Antennas \\
E. Environment-Aware Pinching-Antenna System Designs & F. Pinching-Antenna-Enabled Extremely Large-Scale Antenna Array \\
G. Pinching Antenna-Assisted Wireless Edge Intelligence & H. Robust Pinching-Antenna System Design \\
I. Real-Time Low-Complexity Control of Mechanical Antenna Placement & J. Practical Implementation Challenges and Scalability\\
\multicolumn{2}{c}{\textbf{Section VI. Conclusion}}\\
\hline
\hline
\end{tabular}
\label{tab:outline_comst_colored}
\end{table*}

\subsection{Contributions and Organization of the Paper}
There has been a surge of research interest in pinching-antenna systems recently. Existing studies, including several technical, magazine, and tutorial papers on dielectric waveguide-based implementations, see \cite{ding2025flexible,yang2025pinching,liu2025pinching,liu2025pinching-tutorial} and the references therein, have significantly advanced the understanding and design of such systems. However, as summarized in Table~\ref{tab:comparison_previous_works}, these works mainly focus on a single physical realization, i.e., dielectric waveguides, and address selected aspects in isolation, without offering a unified or systematic treatment that extends the pinching-antenna concept to other realizations or systematically examines the impact of key system parameters on system performance. 

This paper is the first to formally introduce the concept of generalized pinching-antenna systems. This extended framework preserves the core idea of creating localized radiating points on demand but encompasses a variety of physical implementations, including dielectric waveguides, LCXs, and surface-wave guiding structures, together with diverse feeding methods and activation mechanisms. We provide a comprehensive tutorial that covers their fundamental principles, representative architectures, design strategies, integration with emerging wireless technologies, and open research challenges. 
The main contributions of this paper are as follows:
\begin{itemize}
    \item We present a comprehensive overview of the underlying physical mechanisms of generalized pinching-antenna systems, with particular emphasis on their distinctive signal propagation characteristics and reconfigurability. Based on these features, we highlight their unique advantages over conventional-antenna systems.
    \item Using DOCOMO's pinching antennas as an example, we review several representative system architectures, including single- and multi-waveguide configurations. We further survey advanced design strategies for radio resource allocation, beamforming, and antenna placement tailored to pinching-antenna deployments. In addition, we highlight specific system settings where the design problem admits closed-form analytical solutions, which not only enable low-complexity designs but also provide fundamental insights into pinching-antenna systems.
    \item We place a special emphasis on modeling and analyzing the impact of key system parameters, such as in-waveguide attenuation and LoS blockage coefficients. We summarize recent analytical results and present practical guidelines that identify when in-waveguide attenuation can be safely neglected without causing significant performance degradation, under both LoS-dominant and LoS-blockage environments.
    \item We provide an in-depth discussion of emerging applications enabled by pinching antennas, including their integration with existing fixed antennas as well as their use in facilitating ISAC, physical-layer security, wireless power transfer, and other advanced functions. These applications demonstrate the compatibility and practical advantages of pinching antennas, particularly their capability to support dynamic, user-centric, and spatially adaptive wireless services.
    \item We outline several promising research directions and open challenges for future research on pinching-antenna systems. These include pinching-antenna system design under LoS blockage, the realization and optimization of efficient and controllable pinching-antenna radiating points, and the development of accurate and credible electromagnetic models, the exploitation of near-field communication, system optimization under user position uncertainty, and scalable and real-time control strategies.
\end{itemize}

\begin{table*}[ht]
\centering
\caption{List of acronyms.}
\renewcommand{\arraystretch}{1.4}
\begin{tabular}{%
    >{\raggedright\arraybackslash}p{2.0cm} 
    >{\raggedright\arraybackslash}p{5.4cm} 
    >{\raggedright\arraybackslash}p{2.0cm} 
    >{\raggedright\arraybackslash}p{6.2cm}}
\hline\hline
\textbf{Acronym} & \textbf{Definition} & \textbf{Acronym} & \textbf{Definition} \\
\hline
4G     & Fourth-generation & MRT    & Maximum ratio transmission \\
5G     & Fifth-generation  & MU-MISO & Multi-user multiple-input single-output \\
6G     & Sixth-generation  & NLoS   & Non-line-of-sight \\
AI     & Artificial intelligence & NOMA   & Non-orthogonal multiple access \\
AR     & Augmented reality & OFDMA  & Orthogonal frequency-division multiple access \\
AWGN   & Additive white Gaussian noise & OMA    & Orthogonal multiple access \\
BCD    & Block coordinate descent & QoS    & Quality-of-service \\
BS     & Base station & RF     & Radio frequency \\
CPU    & Central processing unit & RIS    & Reconfigurable intelligent surface \\
CR     & Cognitive radio & SCA    & Successive convex approximation \\
CRLB   & Cram\'{e}r--Rao lower bound & SCD    & Semi-cooperative deployment \\
DAS    & Distributed antenna system & SD     & Standalone deployment \\
ELAA   & Extremely large-scale antenna array & SDMA   & Space-division multiple access \\
FCD    & Fully-cooperative deployment & SIC    & Successive interference cancellation \\
FDMA   & Frequency-division multiple access & SINR   & Signal-to-interference-plus-noise ratio \\
FP     & Fractional programming & SISO   & Single-input single-output \\
GNN    & Graph neural network & SNR    & Signal-to-noise ratio \\
IoT    & Internet of things & SWIPT  & Simultaneous wireless information and power transfer \\
IRS    & Intelligent reflecting surface & TDMA   & Time-division multiple access \\
ISAC   & Integrated sensing and communication & TEM    & Transverse electromagnetic \\
KKT    & Karush--Kuhn--Tucker & THz    & Terahertz \\
LCX    & Leaky coaxial cable & ULA    & Uniform linear array \\
LoS    & Line-of-sight & VR     & Virtual reality \\
MIMO   & Multiple-input multiple-output & WMMSE  & Weighted minimum mean square error \\
MMSE   & Minimum mean square error & WPT    & Wireless power transfer \\
mmWave & Millimeter wave & WPU    & Waveguide processing unit \\
\hline\hline
\end{tabular}
\label{tab:abbreviations}
\end{table*}

This paper is organized as follows. Section \ref{sec: pinching antenna fundamentals} introduces the fundamental principles of generalized pinching-antenna systems, including their physical realizations, signal propagation mechanisms, and unique channel characteristics. Representative system architectures and design methodologies for pinching-antenna deployments are systematically surveyed in Section \ref{sec: system design}, where both single- and multi-waveguide systems and the corresponding optimization strategies are covered. Section \ref{sec: application} explores different applications of pinching antennas, detailing their roles in ISAC, security, wireless power transfer, edge computing, and other advanced wireless paradigms. Section \ref{sec: future work} outlines several open questions and research directions. Finally, conclusions are drawn in Section \ref{sec: conclusion}. Table \ref{tab:outline_comst_colored} presents the outline of this paper and Table \ref{tab:abbreviations} lists the acronyms used.

{\emph{Notations:}} Column vectors and matrices are denoted by boldfaced lowercase and uppercase letters, e.g., $\xb$ and $\Xb$, respectively.
$\mathbb{C}^{N}$ denotes the set of $N$-dimensional complex vectors.
Superscripts $(\cdot)^\top$ and $(\cdot)^\Hf$ denote the transpose and Hermitian operations, respectively.
$||\xb||^2$ denotes the square of the Euclidean norm of vector ${\xb}$. 
$\exp(\cdot)$ denotes the exponential function.
$\mathbb{P}(\cdot)$ denotes the probability of an event.
$\E_n[\cdot]$ represents the statistical expectation operation with respect to the random variable $n$. 

\section{Fundamentals of Generalized Pinching-Antenna Systems} \label{sec: pinching antenna fundamentals}
The efficient design of pinching-antenna systems requires an understanding of their underlying physical principles and operational mechanisms. 
In this section, we provide a comprehensive overview of the physical fundamentals underpinning generalized pinching-antenna systems, including the dielectric-based pinching-antenna system, the LCX-based pinching-antenna system, and several types of pinching-inspired antenna systems.

\subsection{Dielectric Waveguide-Based Pinching Antennas}
In this subsection, we first present the system architecture and physical implementation of dielectric waveguide-based pinching-antenna systems, including the design of dielectric waveguides and pinching antenna elements. Then, we illustrate how electromagnetic signals propagate within pinching-antenna systems. We present the distinctive channel model that sets pinching-antenna systems apart from the traditional-antenna systems.

\subsubsection{System Architecture and Physical Implementation}
A typical pinching-antenna system has three components: a BS acting as a central processing unit (CPU) to perform signal processing, dielectric waveguides employed as transmission media, and pinching antennas (small dielectric elements) for dynamic antenna deployment.
As illustrated in Fig. \ref{fig: gpa}(a), the dielectric waveguide is typically connected to the BS through a wired medium. Pinching antennas are strategically placed or dynamically activated along the waveguide, enabling on-demand radiation of electromagnetic energy into free space to form localized communication areas. This unique reconfigurability facilitates rapid deployment and flexible coverage, substantially enhancing the operational flexibility compared with conventional-antenna systems.

\begin{figure*}
    \centering
    \includegraphics[width=0.92\linewidth]{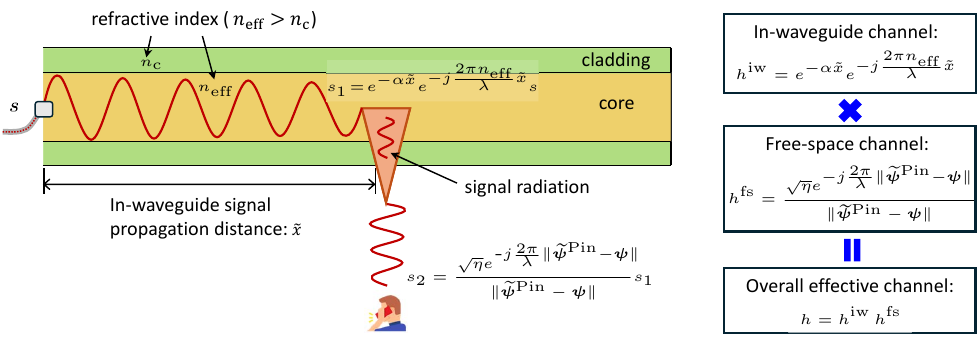}
        \captionsetup{justification=justified, singlelinecheck=false, font=small}	
    \caption{A schematic illustration of signal propagation and channel model for a SISO  pinching-antenna system.}
    \label{fig: pa signal model}
\end{figure*}

A key component of dielectric waveguide-based pinching-antenna systems is a dielectric waveguide, as illustrated in Fig. \ref{fig: pa signal model}. The general structure of a dielectric waveguide includes a waveguiding core made from a dielectric material (such as polytetrafluoroethylene) characterized by a higher refractive index than its surrounding medium, thereby effectively confining and guiding electromagnetic waves within the core. The region with the higher refractive index, denoted by $n_{\textrm{eff}}$, constitutes the \textit{core}, whereas the lower refractive index region ($n_\text{c}$) enveloping the core is referred to as the \textit{cladding} \cite{Kasap13Optoe}. This dielectric waveguide structure exhibits exceptionally low propagation losses at mmWave frequencies, making it suitable for efficient signal transmission over extended distances \cite{snitzer1961cylindrical}.

The operating principle of dielectric waveguide-based pinching antennas involves mechanically attaching discrete dielectric elements—referred to as pinches—to the pre-installed dielectric waveguide. These pinches introduce controlled disruptions to the electromagnetic field within the waveguide, causing localized leakage, i.e., the radiation of electromagnetic waves at the pinch points \cite{kogelnik19752,jackson2012leaky}. Consequently, each pinch functions as a localized radiating antenna capable of both transmitting and receiving signals. The degree and directivity of radiation depend on the geometry, spacing, and dielectric characteristics of these pinching elements relative to the waveguide.

\subsubsection{Signal Propagation and Channel Model}

To illustrate the signal propagation characteristics and derive the corresponding channel model for pinching-antenna systems, we consider a basic single-input single-output (SISO) scenario. In this case, a single pinching antenna mounted on a dielectric waveguide serves one single-antenna user in the downlink, as depicted in Fig.~\ref{fig: pa signal model}. Without loss of generality, we denote the coordinates of the waveguide's feed point as $\bm{\psi}_0 = [0,0,d_v]$, where $d_v$ is the height of the waveguide. The locations of the pinching antenna and the user are given by $\bm{\widetilde \psi}^{\textrm{Pin}} =[\tilde x, 0, d_v]$ and $\bm{\psi} = [\bar x, \bar y, 0]$, respectively.

As illustrated in Fig.~\ref{fig: pa signal model}, the signal propagation process within a pinching-antenna system has two stages:

\begin{itemize}
\item \textit{Stage 1 (In-waveguide propagation):}
The transmit signal $s$ is first fed into the waveguide at the feed point, after which it propagates through the waveguide's core to the position of the pinching antenna. Due to the inherent characteristics of the waveguide, the signal experiences both attenuation and a phase shift. Specifically, the signal reaching the pinching antenna can be represented as
\begin{align}
    s_1 = h^{\textrm{iw}} s = e^{-\alpha \tilde x} e^{-j \frac{2 \pi n_{\textrm{eff}}}{\lambda} \tilde x} s,
\end{align}
where $h^{\textrm{iw}}$ denotes the in-waveguide propagation channel gain (see Fig. \ref{fig: pa signal model}), $n_{\textrm{eff}}$ represents the effective refractive index of the dielectric waveguide, $\lambda$ is the free-space wavelength, and $\alpha$ denotes the in-waveguide attenuation coefficient\footnote{The in-waveguide attenuation coefficient depends on several factors, including the waveguide material, structural design, manufacturing process, and operating frequency. For instance, as reported in \cite{yeh2000communication}, a ceramic alumina ribbon dielectric waveguide designed for mmWave communication can exhibit an in-waveguide attenuation coefficient smaller than $0.1$\,dB/m. Here, we include in-waveguide attenuation in the channel model for notational completeness; however, as will be discussed in Section~\ref{sec: system design}, its impact on system performance is generally negligible under mild and practically achievable system parameter settings.}.

\item \textit{Stage 2 (Free-space propagation):}  
Upon reaching the pinching antenna, the signal $s_1$ is radiated into free space towards the user. That is, the radiated signal then propagates to the user via a free-space channel, characterized by both large-scale path loss and an additional phase shift. By applying the spherical wave channel model \cite{zhang2022beam}, the received signal (by ignoring noise) can thus be expressed as
\begin{align}
    s_2 = h^{\textrm{fs}} s_1 = \frac{\sqrt{\eta} e^{-j \frac{2 \pi}{\lambda}\|  \bm{\widetilde \psi}^{\textrm{Pin}} - \bm{\psi} \|}}{\| \bm{\widetilde \psi}^{\textrm{Pin}} - \bm{\psi} \|}  s_1,
\end{align}
where $h^{\textrm{fs}}$ denotes the free-space channel gain (see Fig. \ref{fig: pa signal model}), and $\eta=\frac{c^2}{(4\pi f_c)^2}$ comes from the Friis equation and captures the frequency dependence of the free-space propagation loss \cite{Balanis16antenna}. Here, $c$ denotes the speed of light and $f_c$ represents the carrier frequency.
\end{itemize}

By concatenating the channels of these two stages, the overall effective channel gain from the waveguide feed point to the user can be modeled as
   \begin{align}\label{eqn: channel model siso}
h = & \ h^{\textrm{iw}}  h^{\textrm{fs}} \notag\\
= & \underbrace{\frac{\eta^{\frac{1}{2}}}{\|\boldsymbol{\widetilde{\psi}}^{\pin} - \bm{\psi}\|}}_{\text{free-space path loss}}   
\underbrace{\exp(-\alpha \tilde x)}_{\text{in-waveguide attenuation}} \notag\\
& 
\exp \bigg({-j \bigg( \underbrace{\tfrac{2\pi}{\lambda} \|\boldsymbol{\widetilde{\psi}}^{\pin} - \bm{\psi}\| + \tfrac{2\pi}{\lambda_g} \tilde x}_{\text{overall phase shift}} \bigg)}\bigg),
\end{align} 
where $\lambda_g = \frac{\lambda}{n_{\textrm{eff}}}$ denotes the guided wavelength within the waveguide.

The channel model derived above highlights several distinctive properties of dielectric waveguide-based pinching-antenna systems. First, unlike the traditional antenna systems, where the antenna positions are fixed, the flexibility to dynamically position pinching antennas along the waveguide enables fine-grained control over path loss. By strategically adjusting antenna placement, the system can reduce propagation distances and establish strong LoS connections, thereby improving the received signal strength. Second, the overall channel gain in pinching-antenna systems captures the phase shifts resulting from both in-waveguide and free-space propagation. This dual phase accumulation, which is absent in conventional wireless channels, means that the channel phase can also be reconfigured through the mechanical repositioning of the antenna or by exploiting the waveguide's propagation characteristics. Such tunability is not available in conventional fixed-antenna systems.
Altogether, these features provide pinching-antenna systems with better adaptability in channel shaping and user-centric link creation. At the same time, they introduce new dimensions of complexity, presenting new challenges and opportunities for the development of optimization algorithms for efficient pinching-antenna system design.

\subsection{LCX-Based Pinching Antennas}
LCX, also known as radiating cable, provides a practical and user-centric means to realize flexible antenna systems, making it a natural extension of the pinching-antenna concept. As shown in Fig. \ref{fig: gpa}(b), in LCX-based systems, the coaxial cable has periodic slots or apertures, which enable controlled and distributed electromagnetic energy leakage at specific locations \cite{torrance1996multi,wang2001theory}.  This allows the cable to function as a long, linear antenna array that supports seamless wireless coverage—an approach widely adopted in underground tunnels, subway systems, and other challenging environments \cite{delogne2003underground,wu2017experimental}.

The core operating principle of LCX is to control the electromagnetic leakage. Structurally, an LCX consists of a central conductor, dielectric insulator, and a slotted or perforated outer shield. As RF signals propagate along the inner conductor, the slotted outer shield guides the energy while allowing a controlled fraction to radiate outward through the slots. The size, spacing, and geometry of these slots, together with the operating frequency, determine the emitted field strength and the uniformity of coverage \cite{wang2016modeling,wu2016performance}. This distributed leakage enables the LCX to act as a long, distributed antenna capable of both transmitting and receiving signals. Due to flexible slot design, LCXs support a broad range of frequencies and coverage requirements, making them especially effective for continuous wireless service in environments where conventional antennas may be less effective.

Recall that the use of waveguide-based pinching antennas ennables flexible on-demand antenna placement, a feature which is crucial to create strong LoS links between the transceivers. Conventional LCX lacks such a flexibility, since the leakage of the signal is due to the presence of slots. These slots are pre-deployed prior to communications, which means that the locations of antennas cannot be tailored to the users' positions in a real-time manner. Recent technological advances have further enhanced LCX's flexibility by enabling the cable to be segmented, with each section equipped with electronic switches or variable couplers \cite{buendia2018smart}. This enables selective activation or deactivation of cable segments, so that only the desired regions are being covered. For example, positive-intrinsic-negative diode or micro-electro-mechanical systems switches can be used to dynamically turn segments on or off in response to user demand or network requirements, creating a reconfigurable and user-centric coverage pattern. Directional couplers and switched taps provide even finer control of radiated fields.

By using the aforementioned technologies, LCX-based systems can also support localized, scenario-driven coverage adjustments, similar to dielectric waveguide-based pinching antennas. However, unlike DOCOMO's pinching antennas, which employ discrete dielectric attachments, LCX-based systems realize this flexibility through controlled leakage slots and segment switching. LCX-based pinching antennas inherit the robustness, low cost, and reliability of coaxial cable technology while offering new flexibility for next-generation wireless networks.

\subsection{Pinching-Inspired Antenna Systems}
The key idea of pinching antennas, realizing on-demand distributed radiation along conductive or engineered surfaces, can be used to design various pinching-antenna-inspired antenna array systems.
In particular, signals are transmitted/received through two stages in both waveguide-based and LCX-based pinching-antenna systems. The first stage is essentially wired transmission. For example, in DOCOMO's pinching-antenna systems, signals are passed through a dielectric waveguide. The second stage is wireless transmission. For example, in DOCOMO's pinching-antenna system, signals are radiated by pinches and sent to the users wirelessly. Therefore, when designing pinching-inspired antenna systems, one question is whether there are other possible signal propagation media in addition to dielectric waveguides and LCX. Ideally, such media should be readily available in practical communication environments and incur a lower cost than dielectric waveguides.

One promising example for the design of pinching-inspired antenna systems is shown in Fig.~\ref{fig: gpa}(c), which is motivated by the following fact. In practical IoT scenarios, there are many metal wires and pipes, such as water and chimney pipes. Through surface-wave communications, these metal pipes can deliver signals to nearby users. Unlike the guided model used by dielectric-waveguide-based transmission, the signals are passed along a metal pipe in the so-called surface-wave mode, i.e., the signals are trapped and propagated on the surface of the pipe \cite{wong2020vision}. However, similar to dielectric-waveguide-based pinching antennas, the first phase of transmission is still wired propagation. Compared with dielectric-waveguide-based pinching-antenna systems, in which coverage is constrained by the length and deployment of dedicated waveguides, such pinching-inspired generalized pinching-antenna systems can exploit existing low-cost media (e.g., metal pipes and wires) and are therefore particularly attractive for extending pinching-antenna concepts to outdoor environments and providing throughput enhancement.
However, this new type of pinching-antenna system based on the surface-wave mode faces two challenges. First, signals propagated via the surface-wave mode may experience higher loss and interference than signals propagated via the guided mode of dielectric waveguides. The second challenge is how to flexibly realize the second stage of pinching-antenna systems, i.e., how to dynamically radiate signals to the user. As shown in Fig. \ref{fig: gpa}(c), one approach is to use active antennas instead of passive antennas, as in waveguide/LCX-based systems. Consider an IoT communication scenario in which a user uses augmented reality (AR) or virtual reality (VR) for smart factory–related training. Conventionally, the serving access point is fixed at the center of the factory and cannot provide a strong LoS link to this user. With the pinching-inspired antenna system, the user can manually attach an active antenna to the surface where a strong LoS link is available and remove the antenna after the training session. 
 
In addition to surface-wave-based pinching antennas, radio-stripe-based pinching antennas can also be designed. For example, electromagnetic signals are first propagated through the radio stripe close to the served user, where the first stage of pinching-antenna transmission, i.e., wired communication, is realized. Then, by incorporating elements such as patched antennas, programmable switches, or directional couplers, a pinching-antenna-inspired architecture can be utilized to provide location-specific, dynamic activation of radiating points, i.e., the second stage of pinching-antenna transmission. While the physical structures of these pinching-inspired antenna systems differ from dielectric waveguides or coaxial cables, the underlying goal remains the same: delivering flexible, spatially controllable, and user-centric wireless coverage with minimal infrastructure overhead.

\begin{table*}[!t]
\centering
\caption{Comparison of representative generalized pinching-antenna systems}
\label{tab:pinching_comparison_vertical}
\renewcommand{\arraystretch}{1.8}
\begin{tabular}{
  >{\centering\arraybackslash}p{3.5cm}
  >{\centering\arraybackslash}p{3.8cm}
  >{\centering\arraybackslash}p{4.2cm}
  >{\centering\arraybackslash}p{4.9cm}}
\hline \hline
\multirow{1.6}{*}{\textbf{Aspect}} &
\textbf{Dielectric-Waveguide-Based Pinching-Antenna System} &
\textbf{LCX-Based Pinching-Antenna System} &
\multirow{1.6}{*}{\textbf{Pinching-Inspired Antenna System}} \\
\hline
\textbf{Signal propagation medium} &
Dielectric waveguide &
Coaxial cable with engineered slots &
Surface of conductor (e.g., metal pipes) \\
\textbf{Propagation mode} &
Guided mode &
Quasi-TEM mode &
Surface-wave mode \\
\textbf{In-medium signal loss} &
Low &
Medium &
High \\
\textbf{Operating frequency} &
High &
Low &
Low \& High \\
\textbf{Active antenna required?} &
No &
No &
Yes \\
\textbf{Deployment scenario} &
Indoor \& Semi-Outdoor &
Indoor \& Outdoor &
Indoor \& Outdoor \\
\textbf{Medium cost} &
Medium &
Medium &
Low \\
\multirow{1.6}{*}{\textbf{Reconfiguration capability}} &
High (continuous position control along waveguide) &
\multirow{1.6}{*}{High (via slot/section configuration)} &
High (continuous position control along metal surface) \\
\hline\hline
\end{tabular}
\end{table*}

Table \ref{tab:pinching_comparison_vertical} provides a comprehensive comparison of three representative generalized pinching-antenna systems, highlighting how each system leverages unique physical and signal propagation characteristics to achieve flexible, on-demand wireless coverage. The ``Propagation Mode'' row describes the fundamental electromagnetic mode by which signals are transported within the medium. For dielectric-waveguide-based pinching-antenna systems, signals are conveyed using the guided mode, which confines the electromagnetic wave within the dielectric structure, resulting in low in-medium signal loss and enabling high-frequency operation. In contrast, LCX-based systems utilize a quasi transverse electromagnetic (TEM) mode characteristic of coaxial cables with engineered slots; this mode provides controlled leakage for distributed coverage but incurs higher signal loss and is typically used at lower frequencies. Pinching-inspired antenna systems, built on metal pipes or radio stripes, exploit surface-wave modes in which the signals propagate along the surface of the medium and then radiate into free space at the desired position. These systems support both low- and high-frequency operation but experience higher in-medium attenuation. These systems also differ in their typical deployment scenarios. Specifically, dielectric-waveguide-based pinching-antenna systems are most suitable for indoor and semi-outdoor environments (e.g., corridors, shopping malls, factories, train stations, or stadium concourses), where waveguides can be conveniently installed and are partly shielded from harsh weather conditions. In contrast, LCX-based and pinching-inspired systems, thanks to their robust mechanical structures and compatibility with existing cables or metal infrastructure, are well suited for both indoor and outdoor deployments. In Table \ref{tab:pinching_comparison_vertical}, the ``Reconfiguration Capability'' row summarizes how the radiation points can be adjusted for each realization: dielectric-waveguide systems naturally support fine, continuous position control along the waveguide, LCX-based systems enable antenna position adjustment via slot or section configuration, and pinching-inspired systems can activate or mechanically tune individual radiating elements on existing conductors.
Overall, this comparison underscores the versatility and adaptability of the generalized pinching-antenna framework, showing how diverse physical mechanisms and design choices enable tailored wireless solutions for a wide variety of coverage needs and environments.

\begin{figure*}[!t]
	\centering
	\includegraphics[width=0.99\linewidth]{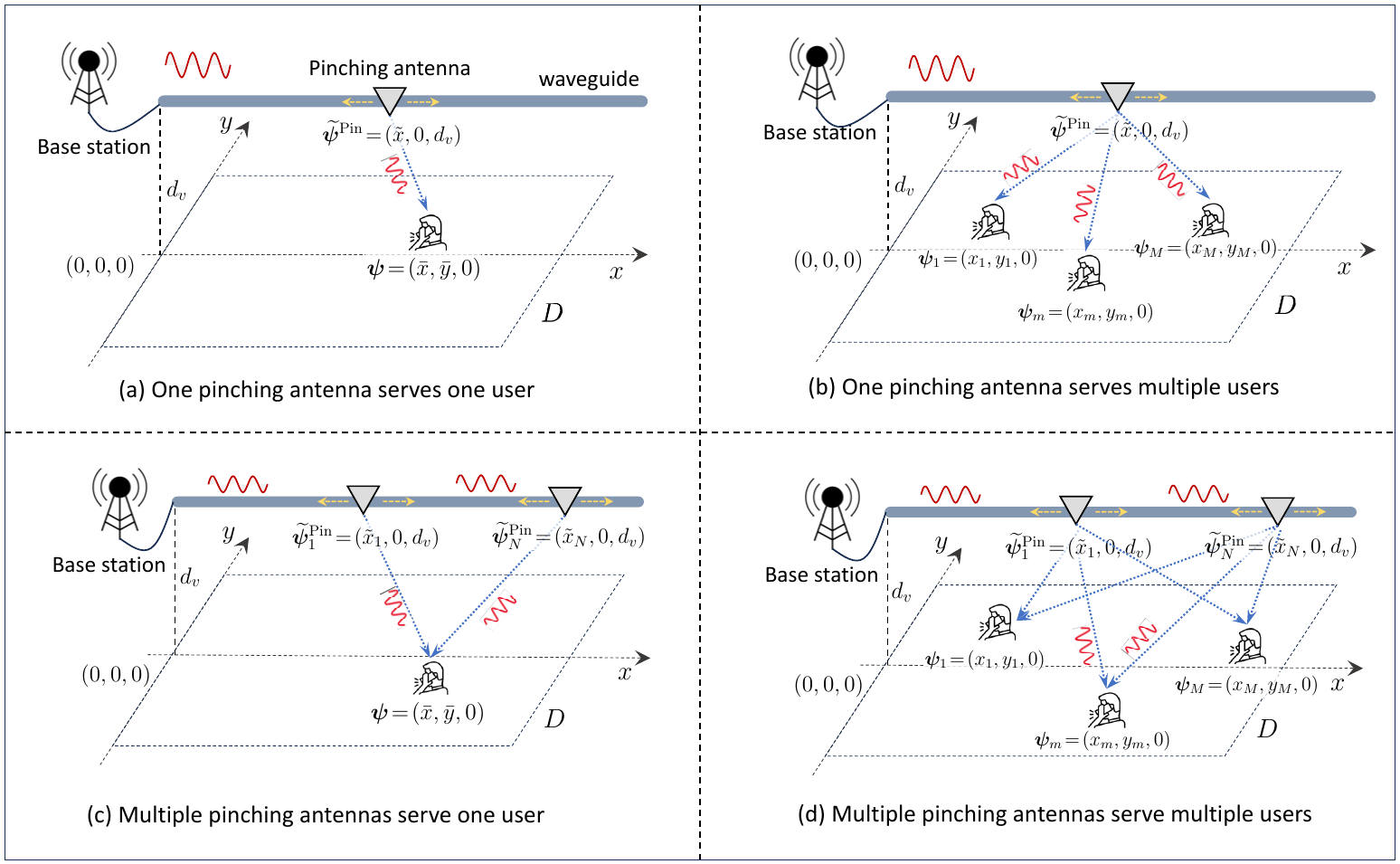}\\
        \captionsetup{justification=justified, singlelinecheck=false, font=small}	
        \caption{Pinching-antenna systems with single waveguide.} \label{fig: system model one waveguide} 
\end{figure*} 

\section{Representative System Architectures and Design Strategies for Pinching-Antenna Systems} \label{sec: system design}

In the remainder of this tutorial, we investigate the impact of generalized pinching antennas on the design of communication systems. Our focus is on the communication network perspective instead of the antenna design perspective. Waveguide-based pinching antennas will be used as the representative example of generalized pinching-antenna systems. The inherent architectural flexibility of pinching-antenna systems paves the way for a wide range of system configurations and optimization strategies tailored for next-generation wireless networks. The resulting system performance and complexity critically depend on key design factors, such as the number of waveguides, the number of pinching antennas, and the employed transmission schemes. Each combination of system parameters and transmission scheme gives rise to distinct optimization problems and challenges. Effectively addressing these interrelated issues is essential for unlocking the full potential of pinching-antenna technology across a broad range of wireless communication scenarios.

We first focus on single-waveguide scenarios, discussing both orthogonal multiple access (OMA\footnote{In this paper, we focus on time division multiple access (TDMA) as an example of OMA for serving multiple users. Alternatively, frequency division multiple access (FDMA) and orthogonal frequency-division multiple access (OFDMA) can also be applied \cite{oikonomou2025ofdma,xiao2025ofdm}.}) and NOMA-based transmission schemes, and detailing the associated optimization methods for antenna placement and resource allocation. Next, we explore multi-waveguide scenarios, where spatial multiplexing, beamforming, and more complex coordination are essential to maximize system performance. 


\subsection{Single-Waveguide Systems}
Pinching-antenna systems having a single dielectric waveguide represent the most basic yet versatile architecture for channel reconfiguration in wireless networks. Depending on the number of pinching antennas deployed and activated, as well as the user configuration, the system can realize a broad range of functionalities and performance trade-offs, as shown in Fig. \ref{fig: system model one waveguide}. 
In this subsection, we discuss two scenarios where a waveguide is equipped with a single pinching antenna and multiple pinching antennas, respectively. For each scenario, we consider both single-user and multiple-user transmissions.

\subsubsection{Single Pinching Antenna on the Waveguide} \label{single antenna noma}
To illustrate the basic design principles and advantages of pinching-antenna systems, we begin by considering the scenario where a single pinching antenna is deployed along a waveguide. This architecture provides a baseline for understanding how antenna placement and user scheduling can be optimized in both single-user and multi-user pinching-antenna systems. Given its simplicity, a pinching-antenna system with a single waveguide and a single antenna is likely to be the first choice in real-world applications. We first focus on the single-user case, highlighting how the flexibility of the pinching antenna can be exploited to improve the system performance. Subsequently, we extend the discussion to multi-user scenarios, examining how OMA and NOMA schemes can further leverage the reconfigurability of the pinching antenna to enhance system performance. This stepwise analysis lays the foundation for addressing more advanced configurations involving multiple pinching antennas or multi-waveguide systems.

\paragraph{Single-User Case}
In this basic case, a single pinching antenna is installed along the waveguide to serve a single-antenna user, as shown in Fig. \ref{fig: system model one waveguide}(a). Without loss of generality, we assume that the user is uniformly distributed over a square communication region with side length $D$, and its location is denoted by $\psib = [\bar x, \bar y,0]$, where $0 \leq \bar x \leq D$ and $-\frac{D}{2} \leq \bar y \leq \frac{D}{2}$.
The waveguide is deployed at the center of this region, parallel to the $x$-axis, with a fixed height $d_v$. The coordinates of the feed point of the waveguide are $\bm{\psi}_0 = [0,0,d_v]$. 
The location of the pinching antenna is given by $\widetilde\psib^{\pin} = [\tilde x, 0, d_v]$.

In this scenario, a typical objective is to maximize the received signal-to-noise ratio (SNR) by optimizing the pinching-antenna position along the waveguide. The problem can be formulated as
\begin{align} \label{p: siso}
    \underset{0 \leq \tilde x \leq D}{\maxn}~ \frac{\rho \eta}{(\tilde x - \bar x)^2 e^{2\alpha\tilde x} + C e^{2\alpha\tilde x}},
\end{align}
where $C = \bar y^2 + d_v^2$ and $\rho = P_{\max}/\sigma_n^2$, with $P_{\max}$ representing the transmit power of the pinching antenna and $\sigma_n^2$ denoting the additive white Gaussian noise (AWGN) power at the receiver.

The denominator of \eqref{p: siso} captures the fundamental trade-off between in-waveguide attenuation and free-space path loss. Specifically, as the pinching antenna moves further from the waveguide feed point, the exponential waveguide attenuation increases while the free-space path loss decreases first and then increases, creating a position-dependent balance that impacts the effective channel gain, and thus, the received SNR.

According to \cite[Lemma 1]{xu2025pinching}, the globally optimal solution to problem \eqref{p: siso} can be derived in closed-form and is given by
\begin{align}\label{eqn: optimal location siso}
    \tilde x^* = 
    \begin{cases}
        0, & \textrm{if~} C \geq \frac{1}{4\alpha^2 } - \frac{(2\alpha \bar x - 1)^2}{4\alpha^2 }, \\
        \bar{x} + \frac{-1 + \sqrt{1 - 4\alpha^2 C}}{2\alpha}, & \textrm{otherwise}.
    \end{cases}
\end{align}

The solution provided in \eqref{eqn: optimal location siso} indicates that the optimal antenna placement is jointly influenced by the user position, the in-waveguide attenuation and the user-to-waveguide distance. Intuitively, if the user is located close to the waveguide or if the waveguide attenuation is negligible, the optimal antenna position approaches the user's projection point on the waveguide (i.e., $\tilde x^* \rightarrow \bar x$), to minimize the free-space path loss. Conversely, as the user moves further from the waveguide or as the in-waveguide attenuation becomes significant, the optimal antenna position gradually shifts toward the waveguide feed point ($\tilde x^* = 0$). This repositioning strategy effectively realizes a balanced tradeoff between reduced in-waveguide attenuation due to reduced propagation distance inside the waveguide, and increased free-space path loss for maximizing the effective received SNR.

A common simplification adopted in the literature is to ignore the in-waveguide attenuation, assuming its impact on performance to be negligible. The motivation behind this assumption is two-fold. Firstly, it significantly simplifies the channel model, making the mathematical analysis more tractable, thereby facilitating theoretical derivations and closed-form results. Secondly, under certain practical conditions, such as high-quality waveguide materials, in-waveguide attenuation effects indeed become relatively small and have an insignificant impact on the overall performance. Under this simplified assumption, the optimal antenna position aligns with the user's projection onto the waveguide, i.e., $\tilde x^* = \bar x$, thus exclusively minimizing free-space path loss.

Naturally, a critical question arises: 

\begin{tcolorbox}[colback=green!10,colframe=gray!50,sharp corners,boxrule=0.5pt]
    \textit{Under what conditions can the in-waveguide attenuation be safely neglected without causing noticeable performance degradation?}
\end{tcolorbox}

From a data rate maximization perspective, the analysis in \cite{xu2025pinching} reveals that ignoring waveguide attenuation is justified, if certain mild and practically achievable conditions on the system parameters, such as waveguide height, attenuation coefficient, and the size of the communication region, are satisfied. 
Specifically, according to \cite[Proposition 1]{xu2025pinching}, the average data rate loss introduced by neglecting waveguide attenuation can be approximated as
\begin{tcolorbox}
\begin{align} \label{eqn: rate gap}
    \mathbb{E}_{\psib}[\Delta R] \approx \frac{\alpha^2}{\ln 2} \left(\frac{D^2}{12} + d_v^2 \right) \quad {{\textrm{(bps/Hz)}}},
\end{align}
\end{tcolorbox}\noindent
where $\Delta R$ denotes the difference between the achievable data rates with and without consideration of the in-waveguide attenuation.

\begin{figure}[!t]
	\centering
	\includegraphics[width=0.99\linewidth]{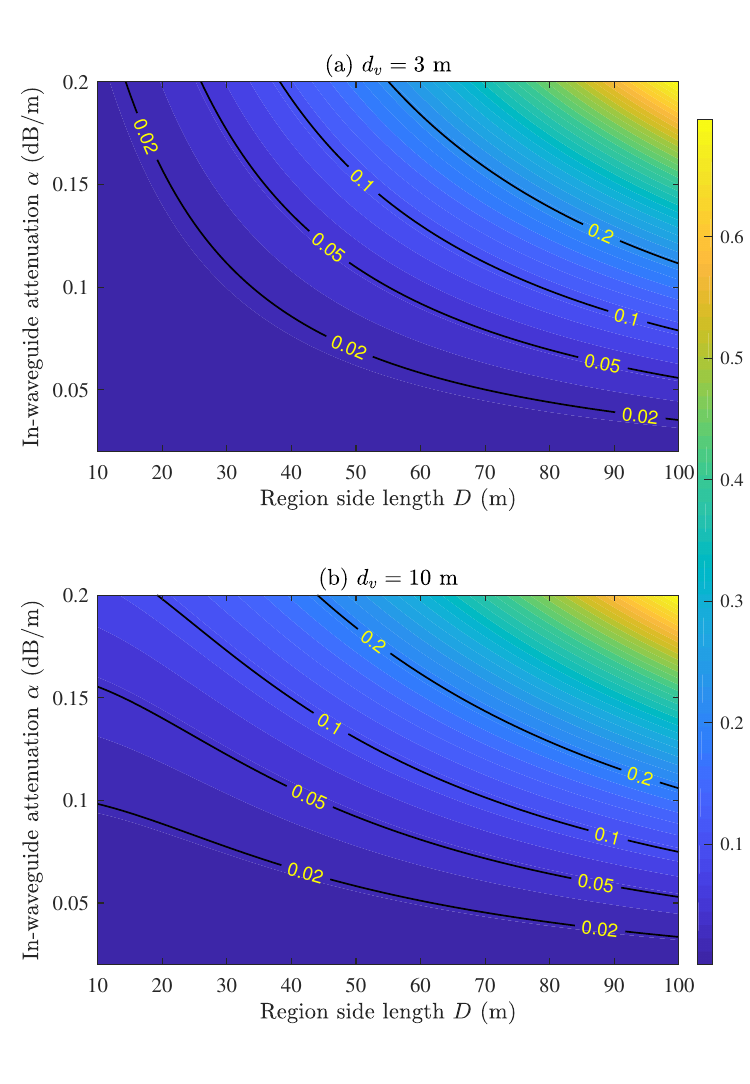}\\
        \captionsetup{justification=justified, singlelinecheck=false, font=small}	
        \caption{Average rate loss $\mathbb{E}_{\psib}[\Delta R]$ versus communication region side length $D$ and in-waveguide attenuation $\alpha$ for (a) $d_v = 3$ m and (b) $d_v = 10$ m.} \label{fig: siso heatmap} 
\end{figure} 

Based on \eqref{eqn: rate gap}, we can directly obtain a simple system design rule: for a given waveguide height $d_v$ and in-waveguide attenuation coefficient $\alpha$, the side length $D$ of the communication region should satisfy 
\begin{align}
    D \leq \sqrt{12 \left(\frac{\epsilon \ln 2}{\alpha^2} - d_v^2\right)},
\end{align}
to ensure that the average rate loss from ignoring in-waveguide attenuation remains below a chosen threshold $\epsilon$. For example, if $d_v = 10$ m, $\alpha = 0.0092$ m$^{-1}$ (corresponds to $0.08$ dB/m)\footnote{
For clarity, in the remainder of this paper, we report the attenuation coefficient $\alpha$ in decibels per meter (dB/m), which is commonly used in experimental studies. Its linear-scale counterpart in nepers per meter (Np/m), often used in the analytical derivations, is given by $\alpha_{\mathrm{Np/m}} = \frac{\alpha_{\mathrm{dB/m}}}{20}\ln(10) \approx 0.1151\,\alpha_{\mathrm{dB/m}}.$ For example, $\alpha = 0.08$\,dB/m corresponds to approximately $0.0092$\,Np/m.}, and the target rate loss is $\epsilon = 0.1$ bps/Hz, we have $D \leq 92.88$ m.
This demonstrates that, for typical system parameters, the region over which the in-waveguide attenuation can be safely neglected is sufficiently large for most practical deployments.

To further illustrate the impact of $(D,\alpha)$ and different waveguide heights, Fig.~\ref{fig: siso heatmap} depicts contour plots of the average rate loss $\mathbb{E}_{\psib}[\Delta R]$ for $d_v = 3$\,m and $d_v = 10$\,m. The contour lines at $0.02$, $0.05$, $0.1$, and $0.2$\,bps/Hz show that, for practical attenuation levels (e.g., $\alpha \le 0.08$\,dB/m) and region sizes $D$ on the order of several tens of meters, the expected rate loss remains below about $0.05$\,bps/Hz so that in–waveguide attenuation can typically be ignored, whereas only when both $\alpha$ and $D$ become simultaneously large (high-loss media and extended coverage regions) does the loss grow appreciably and attenuation-aware design become important.

More recently, the impact of in-waveguide attenuation has been studied in a more general scenario with probabilistic LoS blockage in \cite{xu2025pinching_los}. According to the analysis in \cite[Proposition 1]{xu2025pinching_los}, the average data rate loss due to ignoring in-waveguide attenuation is given by
\begin{tcolorbox}
\begin{small}
\begin{align} 
         &\mathbb{E}_{\psib}[\Delta R]\approx \notag\\
        &\frac{\alpha^2}{\beta \ln 2}\left[ 1 \!-\! \frac{2/D }{\sqrt{\beta(1 \!+\! \beta d_v^2)}} \arctan \left( \frac{\sqrt{\beta}\, D/2}{\sqrt{1 \!+\! \beta d_v^2}} \right) \right],  \quad {{\textrm{(bps/Hz)}}},\label{eq:avg_rate_loss_blockage}
    \end{align}
    \end{small}
\end{tcolorbox}
\noindent
where $\beta \leq 1$ is a small positive constant and denotes the blockage density parameter. It is typically chosen between $0.01$ and $0.1$ m$^{-2}$ \cite{3gpp2020channel} and represents the density of obstacles in the propagation area. A higher $\beta$ value corresponds to a denser blockage environment, thereby reducing the probability of sustaining a LoS connection. The expression in \eqref{eq:avg_rate_loss_blockage} establishes how the average rate degradation is jointly influenced by key system parameters, including the in-waveguide attenuation coefficient $\alpha$, the blockage density parameter $\beta$, the waveguide deployment height $d_v$, and the communication side length $D$. Based on this characterization, several practical insights can be obtained regarding the role of in-waveguide attenuation in pinching-antenna design. Specifically, when $\beta$ is sufficiently small, the expression simplifies to the result in \eqref{eqn: rate gap}, which corresponds to the LoS-dominant scenario. Conversely, in moderate blockage conditions with a sufficiently large communication region (i.e., $\beta D \gg 1$), the average data rate loss $\mathbb{E}_{\psib}[\Delta R]$ converges to the following asymptotic value:
\begin{align} \label{eqn: rate los large D}
    \lim_{D \to \infty} \mathbb{E}_{\psib}[\Delta R]
    = \frac{\alpha^2}{\beta \ln 2},
\end{align}
which implies that as the size of communication area grows (i.e., $\beta D^2 \gg 1$), the performance degradation due to ignoring in-waveguide attenuation approaches a constant and only depends on $\alpha$ and $\beta$. For typical parameters, such as $\alpha = 0.0092$ m$^{-1}$ and $\beta = 0.1$ m$^{-2}$, this constant amounts to merely $0.0012$ bps/Hz. Hence, even in large-scale and blockage-intense environments, the additional loss caused by omitting the in-waveguide attenuation is practically negligible, since the dominant limiting factors are the free-space path loss and LoS blockage in this case.

\begin{figure}[!t]
	\centering
	\includegraphics[width=0.92\linewidth]{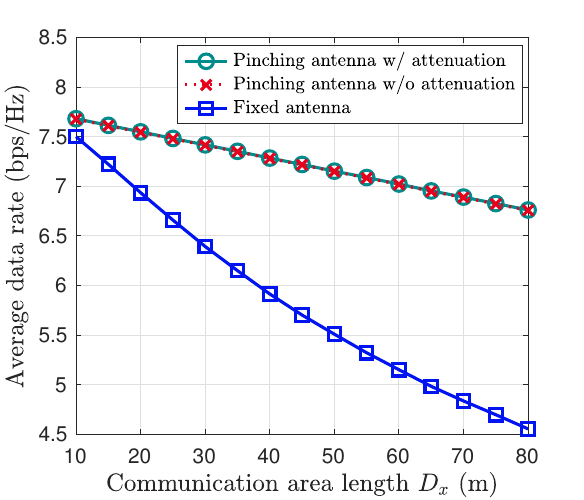}\\
        \captionsetup{justification=justified, singlelinecheck=false, font=small}	
        \caption{Average data rate of pinching-antenna system versus fixed-antenna system.} \label{fig: data rate siso} 
\end{figure}

Accordingly, to streamline both the theoretical development and to facilitate practical insights, we adopt in the remainder of this paper the simplified channel model, where the in-waveguide attenuation is neglected. This modeling choice not only facilitates simpler analytical results but also aligns with typical deployment scenarios where the impact of waveguide attenuation is insignificant.

To further illustrate the performance advantage of pinching-antenna systems compared to the conventional fixed-antenna system, a numerical simulation of the average data rate is presented in Fig. \ref{fig: data rate siso}. In this example, a rectangular communication area is considered with width $D_y = 10$ m and length $D_x$ varying from $10$ to $80$ m, and we assume that there is a LoS channel between the pinching antenna and the user. The waveguide is deployed at the center of this region, parallel to the $x$-axis, with a fixed height $d_v$.
For the pinching-antenna system, two schemes are considered. In the scheme with attenuation, the optimal pinching-antenna position is determined according to \eqref{eqn: optimal location siso}. In contrast, for the scheme without attenuation, the pinching-antenna is simply positioned directly above the user, i.e., $\tilde x = \bar x$.
For the fixed-antenna system, the antenna is fixed at the center of the region, i.e., $[D_x/2,0,d_v]$. The simulation parameters are summarized in Table \ref{tab:sim-param}, which will be reused in the following simulations of this paper unless specified otherwise.

\begin{table}[t]
\centering
\caption{Simulation parameters for Fig. \ref{fig: data rate siso}.}
\renewcommand{\arraystretch}{1.4}
\begin{tabularx}{7.5cm}{>{\raggedright\arraybackslash}p{5.1cm} >{\raggedright\arraybackslash}p{2.5cm}}
\hline \hline
\textbf{Parameter}              & \textbf{Value}           \\
\hline
Carrier frequency ($f_c$)       & $28$ GHz                 \\
Waveguide height ($d_v$)        & $5$ m                    \\
Noise power ($\sigma^2$)        & $-70$ dBm                \\
Transmit power ($P_{\max}$)            & $30$ dBm                 \\
Effective refractive index ($n_{\textrm{eff}}$) & $1.4$              \\
In-waveguide attenuation coefficient ($\alpha$) & $0.0092$ m$^{-1}$ \\
\hline \hline
\end{tabularx}
\label{tab:sim-param}
\end{table}

As shown in Fig.~\ref{fig: data rate siso}, the schemes with and without attenuation achieve nearly identical data rates, indicating that the impact of in-waveguide attenuation is negligible under the given system parameters. This phenomenon has also been numerically validated in \cite{tyrovolas2025performance} from the transmission outage probability perspective.
Additionally, one can observe from Fig.~\ref{fig: data rate siso} that the pinching-antenna system consistently outperforms the fixed-antenna system across all considered region sizes. The performance gain is especially significant as the region size becomes larger, highlighting the advantage of dynamically repositioning the antenna to track the user's location. This result demonstrates the potential of pinching-antenna systems to provide robust, high-rate connectivity particularly in large and dynamic coverage areas, where conventional fixed-antenna systems can suffer from severe path loss for users located at the edge of the communication area.

The main practical and theoretical takeaways from the obtained results are summarized as follows:
\begin{tcolorbox}
    \begin{itemize} \itemsep = 1mm
        \item \textbf{Optimal pinching-antenna position:} When in-waveguide attenuation is small or the user is close to the waveguide, the optimal pinching-antenna position aligns with the user’s projection on the waveguide.
        \item \textbf{Negligible in-waveguide attenuation in practice:} For typical in-waveguide attenuation coefficients and communication region sizes, the impact of in-waveguide attenuation on system performance is often insignificant and can be safely ignored for practical system design.
    \end{itemize}
\end{tcolorbox}

\paragraph{Multiple-User Case} 
Next, let us consider the scenario where $M$ ($M \geq 2$) single-antenna users are served, but only a single pinching antenna is utilized, as illustrated in Fig. \ref{fig: system model one waveguide}(b). In this case, either TDMA or NOMA can be employed. In the following, we discuss each scheme in detail.

\emph{TDMA:}  
In TDMA, each user is allocated an orthogonal time slot for communication \cite{ding2025flexible,samy2025pinching}. 
From an implementation perspective, two pinching antenna configurations can be adopted. In the first configuration, the pinching antenna is dynamically adjusted for each user within their respective time slots. This approach offers the best possible link quality for each user, as the antenna can always be optimally positioned. The resulting optimization is straightforward, since in each slot, the antenna simply aligns with the scheduled user. Alternatively, to reduce system complexity and potential challenges associated with frequent antenna repositioning, a second option is to fix the position of the pinching antenna throughout the entire scheduling period. Hence, the design problem aims to determine the optimal antenna position that serves all users collectively. This approach requires balancing the trade-offs among the different user channel conditions and typically leads to a more challenging optimization problem, necessitating advanced algorithms to identify the best compromise position.

Recently, the authors of \cite{ding2025analytical} provided new analytical insights into antenna placement for pinching-antenna-enabled multi-user communications under OMA transmission. Two practical schemes are investigated: a user-fairness-oriented (max–min) approach and a greedy-allocation-based approach.
In the user-fairness-oriented scheme, the goal is to maximize the minimum achievable rate across all users by jointly optimizing the transmit power and the position of the pinching antenna. The problem is formulated as follows:
\begin{subequations}\label{p: single antenna multiuser tdma}
    \begin{align}
        \underset{P_m \geq 0, \tilde x}{\maxn}~ &\underset{m \in \mathcal{M}}{\rm min.} ~R^{\mathrm{OMA}}_m \\
        \st~&\sum_{m=1}^M P_m \leq P_{\max},\ \ \forall m \in \Mset, \\
        & 0 \leq \tilde x \leq x_{\max}, \label{eqn: x max}
    \end{align}
\end{subequations}
where $\mathcal{M} = \{1, \ldots, M\}$ and $R^{\mathrm{OMA}}_m, \textrm{for}~ m \in \Mset,$ denotes the data rate of user $m$ under OMA transmission. Let $\psib_m = [x_m,y_m,0]$ denote the position of user $m$. In \eqref{eqn: x max}, $x_{\max}$ denotes the maximum side length along the $x$-axis.
Although problem \eqref{p: single antenna multiuser tdma} is non-convex, the analytical results in \cite{ding2025analytical} shown that a closed-form optimal solution can be derived as
\begin{align}
    \tilde x^* &= \frac{1}{M}\sum_{m=1}^M x_m, \\
    P_m &= \frac{\tau_{m,\tilde x^*}}{\sum_{i=1}^M \tau_{i,\tilde x^*}}  P_{\max}, 
\end{align}
where $\tau_{m,\tilde x^*} = (\tilde x^*-x_m)^2 + C_m$ with $C_m = y_m^2 + d_v^2, ~\textrm{for}~m\in \Mset$. This solution reveals that the optimal antenna position is simply the arithmetic mean of all users' $x$-coordinates and—surprisingly—does not depend on the distances between the users and the waveguide. This provides a simple guideline for optimal antenna placement: regardless of the user distances from the waveguide, their x-axis locations alone determine the best antenna position for max–min fairness. The optimal power allocation is proportional to each user's total path loss.

Alternatively, the greedy-allocation-based scheme aims to maximize the system throughput by jointly optimizing the pinching-antenna position and power allocation. The corresponding joint optimization problem is much more challenging than problem \eqref{p: single antenna multiuser tdma} and generally does not admit a closed-form solution. However, the optimal solution to the two-user case under the high-SNR regime can be derived \cite{ding2025analytical}. In particular, results in \cite{ding2025analytical} showed that the optimal pinching-antenna position tends to be closer to the user nearest to the waveguide, in contrast to the fairness-oriented approach. 

These analytical results not only offer a fundamental understanding of pinching antenna placement under OMA but also provide simple, closed-form solutions for optimal deployment in both fairness-driven and throughput-driven use cases.

\emph{NOMA:}  
While OMA offers orthogonal user access and a simple system optimization, it can be spectrally inefficient, especially in scenarios with limited bandwidth resources. Therefore, to improve the spectral efficiency, schemes that can serve all users simultaneously are needed. In this case, NOMA presents a compelling alternative. By allowing multiple users to share the same radio resources and separating their signals in the power domain, NOMA can enhance the spectrum efficiency and user connectivity \cite{ding2017survey,xu2017joint,ding2017application,xu2019transmission,xu2024energy}. This approach is particularly beneficial in pinching-antenna systems where the channel conditions may vary significantly depending on user positions and antenna placement, and there is a strong motivation to serve multiple users concurrently.

For ease of exposition and to illustrate the underlying principles, we first focus on the two-user case. This setting captures the key features and trade-offs of NOMA-assisted pinching-antenna system design while maintaining analytical tractability. 
Without loss of generality, we consider a cognitive radio (CR)-inspired NOMA system \cite{ding2021no,liu2016nonorthogonal}. Specifically, in the considered system, there is a primary user whose quality-of-service (QoS) requirement needs to be satisfied. The other user is an opportunistic secondary user, which seeks to maximize its data rate by utilizing the remaining system resources, after the primary user's needs are satisfied. When NOMA is applied, the signals of both users are combined using superposition coding, i.e., the transmit signal is $s = \sqrt{\alpha_p} s_p + \sqrt{\alpha_s}s_s$, where $s_p$ and $s_s$ denote the signals for the primary user and the secondary user, respectively, and $\alpha_p$ and $\alpha_s$ denote the power allocation coefficients for the primary user and the secondary user, respectively. Here, $\alpha_p$ and $\alpha_s$ are nonnegative and satisfy $\alpha_p + \alpha_s = 1$.
Then, the combined signal $s$ is fed into the waveguide for transmission to the users. After receiving the signal, the primary user directly decodes its intended message, while the secondary user first decodes and subtracts the primary user’s information using successive interference cancellation (SIC) before decoding its own signal. 

For illustration, we consider the problem of maximizing the data rate of the secondary user subject to the QoS requirement of the primary user, by jointly optimizing the locations of pinching antennas and the power allocation coefficients. Let $\psib_m = [x_m, y_m, 0]$ and $\sigma_m^2$ denote the position and received noise power of user $m \in \{p, s\}$, respectively. We define $C_p = y_p^2 + d_v^2$ and $C_s = y_s^2 + d_v^2$. The resulting optimization problem can be formulated as follows:
\begin{subequations} \label{p: rate maximization special 1}
    \begin{align}
        \underset{\alpha_p,\alpha_s, \tilde x}{\rm{maximize}} ~& \frac{\alpha_s P_{\max}\eta}{\big((\tilde x - x_s)^2 + C_s \big)\sigma_s^2} \label{eqn: rate sa obj}\\
        \st ~& \frac{\alpha_p P_{\max} \eta}{\alpha_s P_{\max} \eta + \big((\tilde x - x_p)^2 + C_p\big)\sigma_p^2} \geq \gamma_p, \label{eqn: rate maximization special 1}\\
        & \frac{\alpha_p P_{\max} \eta}{\alpha_s P_{\max} \eta + \big((\tilde x - x_s)^2 + C_s\big)\sigma_p^2} \geq \gamma_p, \label{eqn: rate maximization special 2}\\
        & \alpha_p + \alpha_s = 1, \alpha_p \geq 0, \alpha_s \geq 0,  \label{eqn: rate maximization special 3}
    \end{align}
\end{subequations}
where $\eta = \frac{\lambda^2}{(4\pi)^2}$ is a frequency-dependent constant, with $\lambda$ denoting the free-space wavelength, $P_{\max}$ denotes the transmit power of the pinching antenna and $\gamma_p$ denotes the minimum signal-to-interference-plus-noise ratio (SINR) requirement for the primary user. Constraint \eqref{eqn: rate maximization special 1} corresponds to the data rate requirement of the primary user.
Constraint \eqref{eqn: rate maximization special 2} guarantees the primary user's signal can be successfully decoded at the secondary user by the SIC operation. 

Due to the inherent coupling between the power allocation coefficients and the pinching-antenna position, problem \eqref{p: rate maximization special 1} is non-convex and challenging to solve. Interestingly, as shown in \cite{xu2025qos}, a globally optimal solution to problem \eqref{p: rate maximization special 1} can be derived in closed-form by properly exploiting the structure of the problem. In particular, the optimal solutions of problem \eqref{p: rate maximization special 1} can be obtained by distinguishing between the following three cases:
\begin{itemize}
    \item When $P_{\max}\eta = C_p\sigma_p^2 \gamma_p \geq C_s\sigma_s^2 \gamma_p$, the optimal pinching-antenna position is $\tilde x^* = x_p$, and the optimal power allocation coefficients are $\alpha_p^* = 1$ and $\alpha_s^* = 0$.
    \item When $P_{\max}\eta = C_s\sigma_s^2 \gamma_p \geq C_p\sigma_p^2 \gamma_p$, the optimal pinching-antenna position is $\tilde x^* = x_s$, and the optimal power allocation coefficients are $\alpha_p^* = 1$ and $\alpha_s^* = 0$.
    \item When $P_{\max}\eta > \max\{C_p\sigma_p^2 \gamma_p,C_s\sigma_s^2 \gamma_p\}$, the optimal pinching-antenna position is given by 
    \begin{align} \label{eqn: optimal pinching antenna location 2}
        \tilde x^* = x_p - \beta_p, ~\textrm{or},~ \tilde x^* = \beta_s + x_s,
    \end{align}
    where $\beta_p = \big(\big|\frac{(P_{\max}\eta + P_{\max}\eta\gamma_p) \alpha_p^* - P_{\max}\eta\gamma_p - C_p\sigma_p^2 \gamma_p}{\sigma_p^2\gamma_p}\big|\big)^{\frac{1}{2}}$ and $\beta_s = \big(\big|\frac{(P_{\max}\eta + P_{\max}\eta\gamma_p) \alpha_p^* - P_{\max}\eta\gamma_p - C_s\sigma_s^2 \gamma_p}{\sigma_s^2\gamma_p}\big|\big)^{\frac{1}{2}}$. Besides, the optimal $\alpha_p^*$ is given by 
    \begin{align} \label{eqn: optimal alpha1}
        \alpha_p^* = \max\left\{ \frac{P_{\max}\eta\gamma_p + C_p \sigma_p^2 \gamma_p}{P_{\max}\eta + P_{\max}\eta\gamma_p}, \frac{P_{\max}\eta\gamma_p + C_s \sigma_s^2 \gamma_p}{P_{\max}\eta + P_{\max}\eta\gamma_p} \right\}.
    \end{align}
    and $\alpha_s^* = 1 - \alpha_p^* \in (0,1)$.
\end{itemize}
This closed-form solution reveals several interesting insights for the NOMA-assisted pinching-antenna system design. For example, as the distance between a user and the waveguide increases, the pinching antenna tends to be positioned closer to that user to compensate for the higher path loss. Moreover, when both users are equidistant from the waveguide, the optimal pinching-antenna position is at the midpoint between their projections on the waveguide.

In addition to the rate maximization problem, the transmit power minimization problem has also been investigated in \cite{ding2025analytical} for NOMA transmission with two downlink users. Specifically, the authors of \cite{ding2025analytical} addressed the challenge of jointly optimizing the pinching-antenna position and user power allocation to minimize the total transmit power while ensuring that individual user data rate requirements are satisfied. A closed-form analytical solution is derived, showing that the optimal pinching-antenna position typically favors the user closer to the waveguide. This strategic placement not only enhances the received signal quality for the weaker user but also facilitates efficient SIC. These results provide valuable theoretical insights and offer practical guidelines for deploying NOMA in pinching-antenna systems.

In practice, NOMA-assisted pinching-antenna systems can simultaneously serve more than two users. In such cases, the resulting optimization problem becomes more complex, as it requires jointly determining the optimal pinching-antenna position and the power allocation coefficients for all users, while satisfying individual QoS constraints. The fact that the issues related to SIC, such as SIC decoding order and the success of SIC stages, highly depend on the  pinching-antenna positions makes the multi-user design further challenging. The resulting high-dimensional, coupled optimization problems typically do not admit closed-form solutions. Instead, efficient numerical methods, such as block coordinate descent (BCD), successive convex approximation (SCA), or other iterative methods \cite{xie2025low,zeng2025energy1}, can be employed to find near-optimal solutions in practice. Despite the added complexity, the multi-user NOMA approach can further enhance spectral efficiency and system capacity in dense user environments, making it a promising direction for future pinching-antenna system designs.

\begin{figure}[!t]
	\centering
	\includegraphics[width=0.92\linewidth]{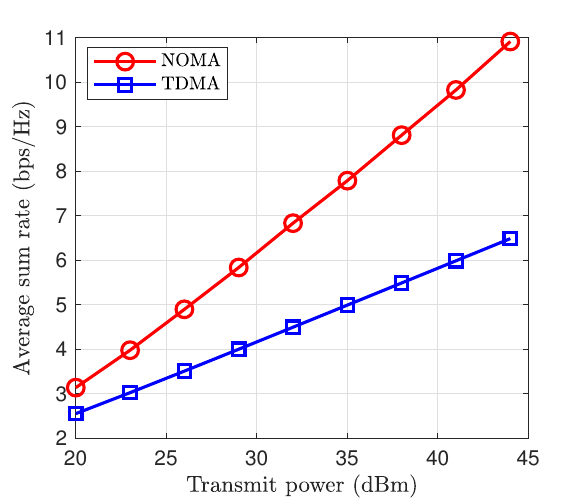}\\
        \captionsetup{justification=justified, singlelinecheck=false, font=small}	
        \caption{Average sum rate of TDMA versus NOMA for a two-user pinching-antenna system with different transmit power.} \label{fig: single antenna noma} 
\end{figure}

{To compare different transmission schemes, Fig.~\ref{fig: single antenna noma} evaluates the average sum rate performance of NOMA and TDMA for a two-user pinching-antenna system. We consider a square communication area with side length equals to $20$\,m, i.e., $D_x = D_y = 20$\,m. The pinching antenna is dynamically positioned according to the design of each scheme. For the TDMA configuration, the transmission frame is divided into two equal-length time slots. Each user is served exclusively in one slot. Within its slot, the pinching-antenna position and transmit power are optimized for that user. For NOMA, both users are served simultaneously in the same time-frequency resource via power-domain multiplexing, with the pinching-antenna position and power allocation jointly optimized. The SNR threshold for the primary user  $\gamma_p$ is set to $1$. The results show that NOMA consistently outperforms TDMA across the considered transmit power range, and the performance gap increases as the transmit power grows. This gain stems from the more efficient use of spectral resources enabled by NOMA, which allows both users to be served simultaneously.

We note that the considered design principles can be extended to scenarios with more than two users by combining pinching-antenna optimization with TDMA/NOMA-based resource allocation.}

The main design principles and takeaways for multiple-user systems are summarized below:

\begin{tcolorbox}
\begin{itemize} \itemsep = 1mm
    \item \textbf{Low-complexity TDMA Scheme:} Repositioning the pinching antenna for each user in its allocated time slot achieves near-optimal performance based on analytical solutions, while a fixed antenna position offers reduced system complexity at the cost of a performance loss.
    \item \textbf{Spectrally-Efficient NOMA Scheme:} NOMA efficiently improves the system spectrum efficiency, where joint optimization of the antenna position and the user power allocation coefficients is crucial for balancing data rates and guaranteeing QoS constraints. 
    \item \textbf{Advanced Optimization Methods:} For scenarios with more than two users or more complex system requirements, computationally intensive iterative algorithms are often needed if NOMA scheme is employed. Developing more efficient algorithms for these challenging optimization problems remains as an important and promising direction for future research.
\end{itemize}
\end{tcolorbox}

\subsubsection{Multiple Pinching Antennas on the Waveguide}
Deploying multiple pinching antennas along a single waveguide unlocks new opportunities for performance enhancement and flexible resource allocation \cite{ding2025flexible,ouyang2025array}. In this subsection, we explore system configurations that allow the simultaneous activation of multiple pinching antennas on a single waveguide. 
It is important to note that, although multiple pinching antennas can be simultaneously deployed along the waveguide, conventional beamforming techniques cannot be directly applied. 
This is because only one signal can be transmitted and propagated through the waveguide at any given time, precluding the possibility of transmitting independent, spatially multiplexed data streams as in traditional MIMO systems \cite{ding2025flexible}. 
Instead, the additional degree of freedoms offered by multiple pinching antennas can be exploited by carefully optimizing the pinching-antenna positions so that the signals emitted from each antenna arrive at the user with aligned phases. This constructive alignment allows the individual signals to combine coherently at the receiver, thus enhancing the overall received signal strength. In this way, the considered pinching-antenna system with multiple antennas is deployed on a single waveguide is similar to analog beamforming, but the phases of the channel gains are controlled not by a phase shifter but by selecting the positions of the pinching antennas. 

In what follows, we first consider the single-user scenario to highlight the advantages of deploying multiple pinching antennas on a waveguide, as well as the associated design challenges \cite{xu2025rate,ouyang2025array}. We then extend the discussion to the multi-user case, where system design must address the joint optimization of the pinching-antenna position and radio resources in order to fully exploit the additional degrees of freedom offered by multiple pinching antennas \cite{xu2025qos,tegos2024minimum,mu2025pinching}.

\paragraph{Single-User Case}
Let us first focus on the single-user case, as illustrated in Fig. \ref{fig: system model one waveguide}(c). 
Denote the user signal by $s$, which is fed into the waveguide and transmited to the user via $N$ pinching antennas. Denote the positions of the user and pinching antenna $n$ by $\psib = [\bar x, \bar y, 0]$ and $\tilde\psib_n^{\pin} = [\tilde x_n, 0, d_v]$, respectively. The channel between pinching antenna $n$ and the user is given by
\begin{align}
    h_n = \frac{\eta^{\frac{1}{2}} e^{-j\big(\frac{2 \pi}{\lambda}[(\tilde x_n - \bar x)^2 + C ] + \frac{2 \pi}{\lambda_g}\tilde x_n\big)}}{\big[(\tilde x_n - \bar x)^2 + C \big]^{-\frac{1}{2}}},
\end{align}
where {$C = \bar y^2 + d_v^2$}.
Without loss of generality, we assume that the pinching antennas are deployed in a successive order, i.e., $\tilde x_n - \tilde x_{n-1} > 0, \forall n \in \Nset \triangleq \{1,...,N\}$.
The objective is to maximize the received SNR by jointly optimizing the positions of all pinching antennas, which leads to the following problem formulation:
\begin{subequations}\label{p: rate maximization multi waveguide single user}
    \begin{align}
     \underset{\tilde x_1,...,\tilde x_n}{\maxn} ~&\bigg| \sum\limits_{n=1}^{N} \frac{ e^{-j \phi_n }}{(\tilde x_n - \bar x)^2 + C} \bigg|\\
     \st ~& \tilde x_{\min} \leq \tilde x_n \leq \tilde x_{\max}, \forall n \in \Nset, \\
     ~& \tilde x_n - \tilde x_{n-1} \geq \Delta, \forall n \in \Nset \setminus \{1\},\\
     ~& \phi_n = \frac{2 \pi}{\lambda}[(\tilde x_n - \bar x)^2 + C ] + \frac{2 \pi}{\lambda_g}\tilde x_n, \forall n \in \Nset,
    \end{align}
\end{subequations}
where $\Delta$ denotes the minimum inter-antenna spacing. 
Problem \eqref{p: rate maximization multi waveguide single user} is challenging to solve because the pinching-antenna positions affect two aspects of the channel. One is the large-scale path loss, and the other one is the phase shift due to signal propagation inside and outside the waveguide. 

To maximize the objective function, one needs to minimize the impact of the large scale path loss, and meanwhile, needs to ensure that the received signals from different pinching antennas are constructively combined at the user. To achieve this goal, one can solve the following approximated problem:
\begin{subequations}\label{p: rate maximization approx 1}
    \begin{align}
     \underset{\tilde x_1,...,\tilde x_n}{\maxn} ~& \sum\limits_{n=1}^{N} \big[(\tilde x_n - x)^2 + C \big]^{-\frac{1}{2}} \label{eqn: obj}\\
     \st ~&\tilde x_n - \tilde x_{n-1} \geq \Delta, \forall n \in \Nset \setminus \{1\},\\
     ~& \phi_n -\phi_{n-1} = 2k\pi, \forall n \in \Nset \setminus \{1\}, \label{eqn: phase constraints}
    \end{align}
\end{subequations}
where $k$ is an arbitrary integer.
Building on this formulation, a low-complexity two-stage optimization algorithm was proposed in~\cite{xu2025rate} to efficiently address the design challenge. In the first stage, the pinching-antenna positions minimizing the total large-scale path loss between the antennas and the user are determined. This step focuses on optimizing the geometric configuration so that each antenna is as close to the user as possible, under the constraints imposed by the antenna spacing. In the second stage, the antenna positions are further fine-tuned to align the phases of the signals from all antennas, ensuring that the signals are combined constructively at the user, thereby maximizing the total received SNR.
This two-stage approach effectively decouples the complicated optimization problem into more manageable subproblems, offering a practical and scalable solution for maximizing the received signal strength in multi-pinching-antenna systems.

\paragraph{Multiple-User Case}
We now turn our focus to the case where multiple users are to be served by multiple pinching antennas deployed along a waveguide, as illustrated in Fig. \ref{fig: system model one waveguide}(d). Without loss of generality, we assume that the number of pinching antennas is $N$ and the number of users is $M$. Denote the positions of pinching antenna $n$ and user $m$ by $\widetilde \psib_{n}^{\pin} = [\tilde x_n, 0, d_v]$ and $\psib_m = [x_m, y_m, 0]$, respectively.
The effective channel between the $N$ pinching antennas and user $m$ is given by
\begin{align} \label{eqn: channel multiple pinching antenna}
    h_m = \sum_{n=1}^N \frac{\eta^{\tfrac{1}{2}} e^{-j\big(\frac{2 \pi}{\lambda}[(\tilde x_n - \bar x)^2 + C_m ] + \frac{2 \pi}{\lambda_g}\tilde x_n\big)}}{\big[(\tilde x_n - \bar x)^2 + C_m \big]^{-\frac{1}{2}}}, \forall m \in \Mset,
\end{align}
where $C_m = y_m^2 + d_v^2$ and $\Mset \triangleq \{1,...,M\}$. 

Similar to the single-antenna setting, both TDMA and NOMA transmission can be employed. In the following, we discuss the design principles, resource allocation strategies, and the associated challenges for each scheme.

\emph{TDMA:} In the multi-user scenario with multiple pinching antennas deployed on a waveguide, TDMA enables each user to be served by all antennas during its allocated time slot. Depending on whether the pinching-antenna positions can be dynamically optimized in each time slot, TDMA can be implemented via the following two schemes. In the first scheme, the pinching-antenna positions are re-optimized in each user's time slot, effectively reducing the problem to a sequence of single-user configurations, where the previously described algorithm for the single-user case can be applied \cite{hou2025performance}. In the second scheme, the pinching-antenna positions are optimized only once for the entire scheduling interval. While this strategy reduces the mechanical complexity and overhead associated with frequently repositioning antennas, it significantly increases the design challenge, as the pinching-antenna positions have to be jointly optimized for all users \cite{tegos2024minimum}. In the following, we provide an example to illustrate the latter scheme.

Denote the time duration allocated to user $m$ by $t_m$. The achievable data rate for user $m$ can be expressed as
\begin{align}
    R_m = t_m \log\left(1 + \frac{P_{\max} |h_m|^2}{N\sigma_m^2 t_m} \right).
\end{align}
To ensure fairness among users, we focus on maximizing the minimum achievable data rate across all users. The resulting max–min fairness problem is formulated as follows:
\begin{subequations} \label{p: max min tdma}
    \begin{align}
        \underset{\tb, \tilde \xb}{\maxn}~ &\underset{m \in \mathcal{M}}{\rm min.} ~R_m \\
        \st ~& \tilde x_{\min} \leq \tilde x_n \leq \tilde x_{\max}, \forall n \in \Nset, \label{eqn: max min tdma 1}\\
        & \tilde x_n - \tilde x_{n-1} \geq \Delta, \forall n \in \Nset \setminus \{1\},\label{eqn: max min tdma 2}\\
        & \sum_{m=1}^M t_m = T, t_m \geq 0, \label{eqn: max min tdma 3}
    \end{align}
\end{subequations}
where $\tb = [t_1,\ldots,t_M]^{\top}$ represents the time allocation vector, $\tilde \xb = [\tilde x_1,\ldots,\tilde x_N]^{\top}$ collects the pinching-antenna positions, and $T$ denotes the total transmission duration.

Owing to the coupling between user scheduling and antenna positioning, problem~\eqref{p: max min tdma} is highly non-convex and challenging to solve directly. To tackle this, an alternating optimization approach can be employed, where the problem is decomposed into two more manageable subproblems:
\begin{itemize}
    \item {\textit{Pinching-antenna position optimization:}} For a fixed time allocation $\tb^*$, the antenna positions $\tilde \xb$ are optimized by solving
    \begin{align*}  
        {\textrm{(P1)}} ~\underset{\tilde \xb}{\maxn}~ &\underset{m \in \mathcal{M}}{\rm min.} ~ R_m(\tb^*,\tilde \xb)\\
        \st ~& \eqref{eqn: max min tdma 1}, \eqref{eqn: max min tdma 2},
    \end{align*}
    which is still non-convex due to the objective function's structure, as the shared antenna positions must be simultaneously tailored to all user locations. A practical approach is to use the SCA method, which iteratively approximates the non-convex parts by convex surrogates, leading to a sequence of tractable convex optimization subproblems \cite{zhou2025gradient,xu2025qos}.
    \item {\textit{Time duration allocation:}} Once the antenna positions $\tilde \xb^*$ are fixed, the optimal time allocation $\tb$ is obtained by solving
    \begin{align*} 
        {\textrm{(P2)}} ~\underset{\tb}{\maxn}~ &\underset{m \in \mathcal{M}}{\rm min.} ~R_m(\tb,\tilde \xb^*) \\
        \st ~& \eqref{eqn: max min tdma 3},
    \end{align*}
    which is convex and can be efficiently solved either using standard convex optimization solvers or in closed-form via the Lagrangian method \cite{tegos2024minimum}.
\end{itemize}
By alternating between these two subproblems, the algorithm iteratively improves both the pinching-antenna positions and the time allocation, converging to a locally optimal solution that balances user fairness and system throughput. This approach offers a practical framework for the challenging resource allocation task in TDMA-based multi-user pinching-antenna systems.

Besides alternating optimization with SCA, a recent study \cite{xie2025graph} has investigated the application of graph neural networks (GNNs) to address these challenges. In this framework, the objective is to jointly optimize the antenna placement and power allocation to maximize the system's energy efficiency, subject to transmit power constraints and minimum inter-antenna spacing. To overcome the high computational complexity faced by conventional optimization methods, the GNN-based method treats the wireless system as a bipartite graph, with user and antenna nodes connected by edges that encode their spatial relationships. A bipartite GNN is then employed to learn effective mappings from the system state—comprising user and antenna positions as well as current channel conditions—to resource allocation decisions, specifically the antenna positions and the power allocated to each antenna. The learning process is supervised by training data generated from solutions obtained through conventional optimization techniques, enabling the GNN to approximate complex decision boundaries without requiring real-time, iterative optimization. Once the model has been trained, the GNN-based method can produce high-quality resource allocation policies with significantly lower computational overhead, making it attractive for practical systems with stringent latency requirements or large-scale user deployments.

\emph{NOMA:} The NOMA scheme seeks to jointly optimize the antenna positions and power allocation coefficients so that the received signals at each user can be effectively decoded \cite{xu2025qos,wang2024antenna,zhou2025sum,hu2025sum,fu2025power,zeng2025sum,zeng2025energy1,gan2025joint,ren2025pinching}. Unlike the single-antenna case in Section \ref{single antenna noma}, where an optimal closed-form solution can be obtained, the problem for the multiple antenna case is much more challenging. The fundamental challenge in the NOMA-based design lies in the strong coupling between the pinching-antenna positions and the power allocation among users. In what follows, we focus on the case of the two-user CR-NOMA scheme, aiming to clearly illustrating the key challenges in solving the optimization problem and representative solutions. 

For this setup, consider the maximization of the achievable data rate of the secondary user while ensuring that the primary user's QoS requirement is strictly satisfied. The associated optimization problem can be formulated as follows:
\begin{subequations} \label{p: rate maximization origin}
    \begin{align}
        \underset{\substack{\alpha_p,\alpha_s, \tilde \xb}}{\rm{maximize}} ~ &\frac{\alpha_s P_{\max} |\tilde h_s|^2}{N\sigma_s^2} \\
        \st ~&  \alpha_p P_{\max} |\tilde h_p|^2 \geq \gamma_p \big(\alpha_s P_{\max} |\tilde h_p|^2 + N\sigma_p^2 \big), \label{eqn: rate maximization origin 1}\\
        & \alpha_p P_{\max} |\tilde h_s|^2 \geq \gamma_p \big(\alpha_s P_{\max} |\tilde h_s|^2 + N\sigma_s^2\big), \label{eqn: rate maximization origin 2}\\
        & \tilde x_n - \tilde x_{n-1} \geq \Delta, \hspace{3mm}\forall n\in \Nset \setminus \{ 1 \}, \label{eqn: rate maximization origin 3}\\
        & \alpha_p + \alpha_s = 1, \alpha_p\geq 0, \alpha_s\geq 0, \label{eqn: rate maximization origin 4}
    \end{align}
\end{subequations}
where $\tilde h_m$ denotes the channel between the pinching antennas and user $m \in \{p,s\}$.
Here, constraint \eqref{eqn: rate maximization origin 1} ensures that the primary user’s data rate meets the required SINR threshold, while constraint \eqref{eqn: rate maximization origin 2} guarantees that the primary user’s signal can be successfully decoded by the secondary user using SIC. The minimum spacing between adjacent pinching antennas is imposed by constraint \eqref{eqn: rate maximization origin 3}, and constraint \eqref{eqn: rate maximization origin 4} corresponds to the constraint on the power allocation coefficients. 

Problem \eqref{p: rate maximization origin} is challenging to solve due to the intricate coupling between the power allocation coefficients and the effective channel gains, which are nonlinear functions of the pinching-antenna positions. To address this, a practical solution is to adopt a BCD approach, which decomposes the original problem into more tractable subproblems that can be solved iteratively \cite{xu2025qos}. Specifically, the optimization proceeds as follows:
\begin{itemize}
    \item \textit{Power allocation optimization:} Given the pinching-antenna positions, the optimal power allocation coefficients $(\alpha_p, \alpha_s)$ can be derived in closed-form based on the analysis in \cite{xu2025qos}.
    Notably, if the optimal pinching-antenna positions are available, e.g., by a linear search, this closed-form solution provides the globally optimal power allocation. This result offers a valuable theoretical benchmark for evaluating practical suboptimal algorithms with low computational complexity. However, exhaustive search quickly becomes infeasible as the number of antennas increases, motivating the need for efficient, scalable algorithms.

    \item \textit{Pinching-antenna position optimization:} With the power allocation fixed, the optimization reduces to optimizing the pinching-antenna positions for maximizing the data rate of the secondary user. To deal with the complexity of the channel gains with respect to the antenna positions, iterative techniques, such as SCA, can be applied to efficiently approximate and solve this subproblem \cite{xu2025qos}.
\end{itemize}
By iteratively solving the two subproblems until a convergence criterion is satisfied, this approach provides a practical means to deal with the challenging joint design of power allocation and antenna positioning in NOMA-assisted pinching-antenna systems.

\begin{figure}[!t]
	\centering
	\includegraphics[width=0.92\linewidth]{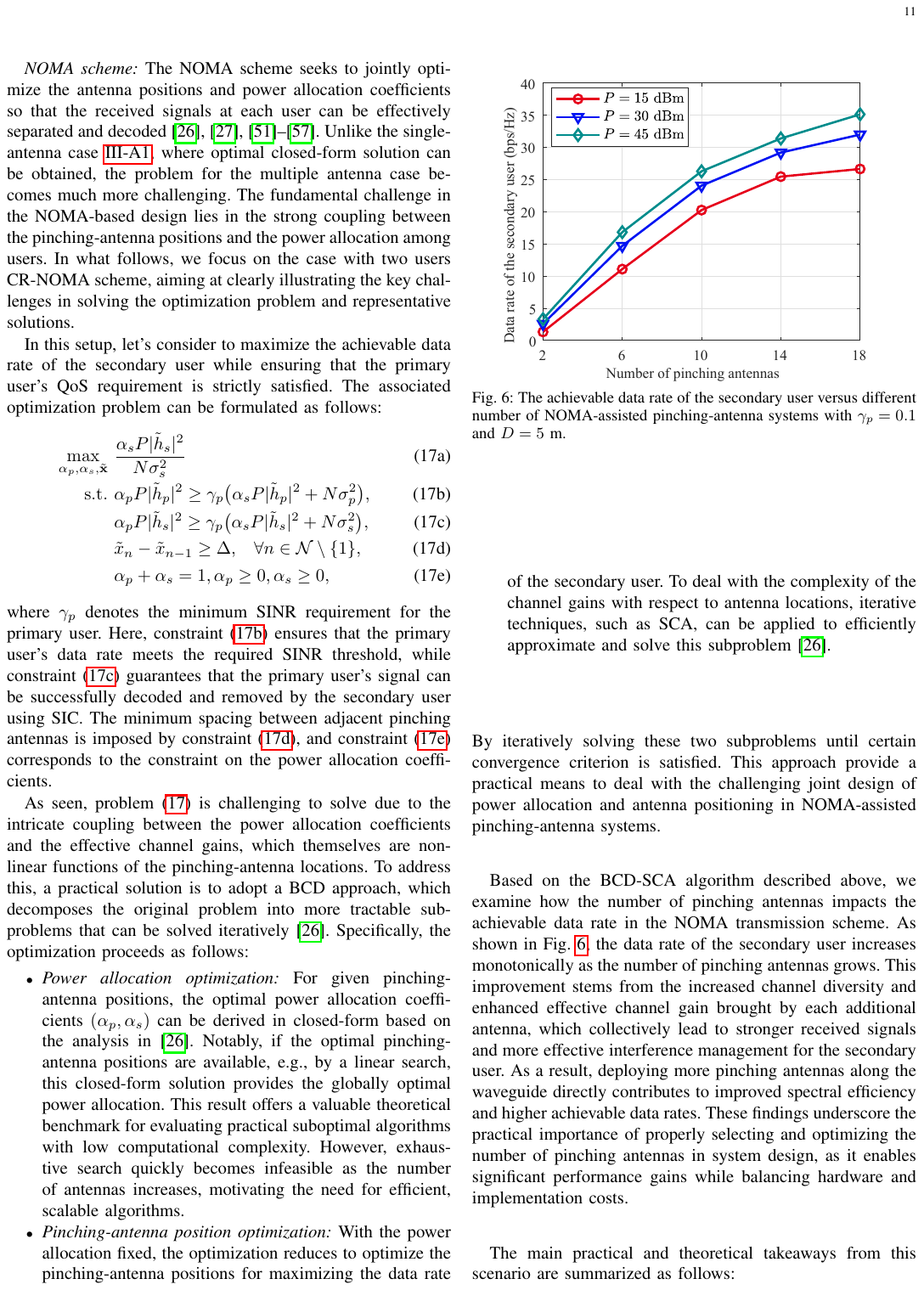}\\
        \captionsetup{justification=justified, singlelinecheck=false, font=small}	
        \caption{Achievable data rate of the secondary user versus different numbers of pinching-antennas with $\gamma_p = 0.1$.} \label{fig: rate n}
\end{figure} 

Based on the BCD-SCA algorithm described above, we examine how the number of pinching antennas impacts the achievable data rate in the NOMA transmission scheme. The results are shown in in Fig. \ref{fig: rate n}, where the parameters in Table \ref{tab:sim-param} are used and the two users are randomly deployed in a square region with side length $D = 5$ m. As seen, the data rate of the secondary user increases monotonically as the number of pinching antennas grows. This improvement stems from the increased channel diversity and enhanced effective channel gain brought by each additional antenna, which collectively lead to stronger received signals and more effective interference management for the secondary user. As a result, deploying more pinching antennas along the waveguide directly contributes to improved spectral efficiency and higher achievable data rates. These findings underscore the practical importance of properly selecting and optimizing the number of pinching antennas in system design, as it enables significant performance gains while balancing hardware and implementation costs.

The main practical and theoretical takeaways from this scenario can be summarized as follows:

\begin{tcolorbox}
\begin{itemize}\itemsep = 1mm
    \item \textbf{Spatial Diversity Gains:} Deploying multiple pinching antennas enables significant spatial diversity and potential array gains, which can be exploited to improve system performance.
    \item \textbf{Constructive Signal Reception:} Unlike conventional MIMO systems, beamforming is not feasible as only one signal can be injected into the waveguide at a time. Instead, optimizing the pinching-antenna positions to ensure that the received signals from all antennas are constructively combined at the user is crucial for performance improvement.
    \item \textbf{System Design Challenge:} In the multi-antenna scenario, the system design problem becomes considerably more complex, as it requires the joint optimization of pinching-antenna positions and resource allocation. Developing scalable, low-complexity algorithms for these highly coupled, non-convex problems, especially for large-scale or real-time systems, remains an important and open research direction.
\end{itemize}
\end{tcolorbox}

\begin{figure*}[!t]
	\centering
	\includegraphics[width=0.99\linewidth]{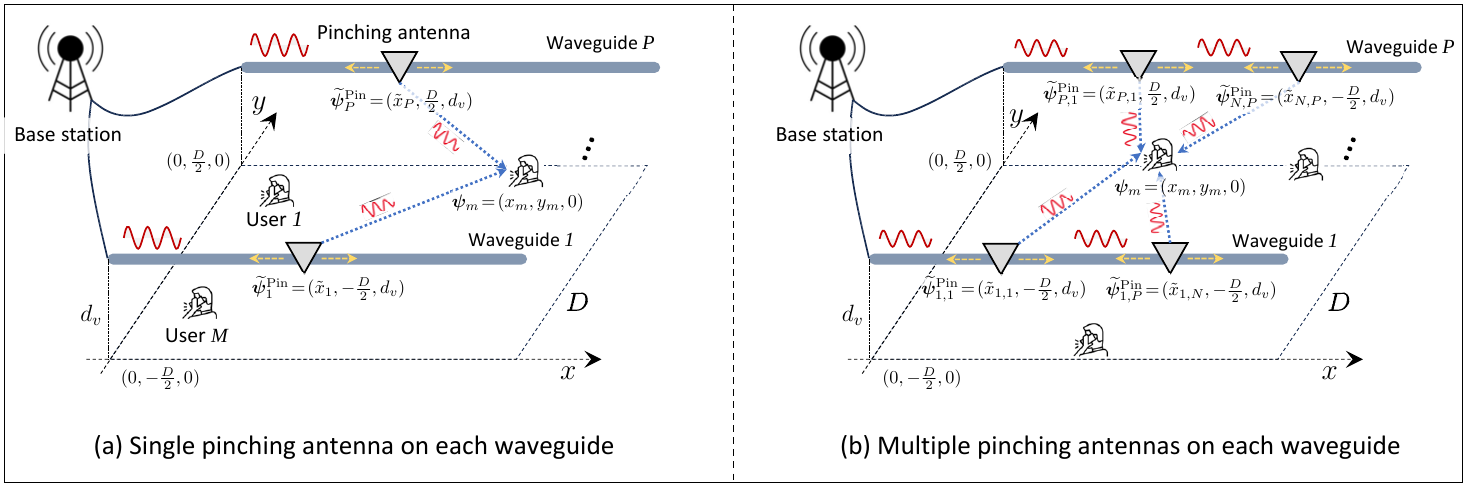}\\
        \captionsetup{justification=justified, singlelinecheck=false, font=small}	
        \caption{Pinching-antenna systems with multiple waveguides.} \label{fig: system model multiple waveguide} 
\end{figure*}

\subsection{Multiple-Waveguide Systems}
\label{sec:Multiple-Waveguide Systems}
To further enhance system flexibility and spatial diversity, multiple waveguides can be deployed in pinching-antenna systems. In this subsection, we systematically discuss design approaches for such multi-waveguide configurations by focusing on two representative scenarios. The first scenario considers that each waveguide is equipped with a single pinching antenna, while the second one explores the use of multiple pinching antennas per waveguide. In the remainder of this subsection, we present the signal models, formulate the corresponding optimization problems, and highlight representative solutions for each case.

\subsubsection{Single Pinching Antenna Per Waveguide} 
We begin by considering the scenario with $M$ single-antenna users and $P$ waveguides, where a single pinching antenna is activated on each waveguide \cite{xu2025pinching,zhang2025uplink,qian2025pinching}. The waveguides are positioned parallel to the $x$-axis at height $d_v$, and the distance between two neighboring waveguides is $d_h = \frac{D}{P-1} \gg \lambda$, as illustrated in Fig. \ref{fig: system model multiple waveguide}(a). 
Without loss of generality, we assume that the coordinate of the feed point of the $p$-th waveguide is $\bm{\psi}_{0,p} = [0,(p-1)d_h - \frac{D}{2},d_v], p\in \mathcal{P} \triangleq \{1,...,P\}$. The position of the pinching antenna on the $p$-th waveguide is denoted by $\widetilde\psib_{p}^{\pin} = [\tilde{x}_p, (p-1) d_h - \frac{D}{2}, d_v]$ and the position of user $m$ is denoted by $\psib_m = [x_m,y_m,d_v]$. 
Then, the channel coefficient between pinching antenna $p$ and user $m$ is given by 
\begin{align} \label{eqn: channel multiple waveguide 1}
    h_{m,p} \!=\! \frac{\eta^{\frac{1}{2}} e^{-j \big( \frac{2\pi}{\lambda} \big[(\tilde x_p - x_m)^2 + C_{m,p}\big]^{\frac{1}{2}} + \frac{2\pi}{\lambda_g} \tilde x_p \big)}} {\big[(\tilde x_p - x_m)^2 + C_{m,p}\big]^{\frac{1}{2}} e^{\alpha\tilde x_p}},  \forall p \!\in \! \mathcal{P},  m \!\in\! \mathcal{M},
\end{align}
where $C_{m,p} = \big((p-1)d_h - \frac{D}{2} - y_m\big)^2 + d_v^2$. 
Then, the sum rate maximization problem can be formulated as follows
\begin{subequations} \label{p: sum rate maximization}
    \begin{align}
        \underset{\tilde \xb, \vb_m, \forall m}{\rm maximize} ~&\sum_{m=1}^M \log_2\left(1 + \frac{|\hb_m^\top \vb_m|^2}{\sum_{i \neq m}|\hb_m^\top \vb_i|^2 + \sigma_m^2}\right)\\
        \st~ & \sum_{m=1}^M \|\vb_{m}\|^2 \leq P_{\max},  \label{eqn: papc}\\
        & 0 \leq \tilde x_n \leq x_{\max}, \forall n \in \Nset,
    \end{align}
\end{subequations}
where $\tilde \xb = [\tilde x_1,...,\tilde x_N]^\top$ collects the locations of the pinching antennas, $\hb_m = [h_{m,1},..., h_{m,P}]^\top$ stores the channel vector between the $P$ pinching antennas and user $m$, and $\vb_m \in \Cs^{N}$ represents the beamformer for user $m$.
Constraint~\eqref{eqn: papc} enforces the total transmit power limit.

It is worth noting that when $M = 1$, i.e., only a single user is served, the optimal beamformer reduces to the classical maximum ratio transmission (MRT) beamformer, and problem~\eqref{p: sum rate maximization} simplifies to optimizing the pinching-antenna positions only. As shown in \cite{xu2025pinching}, the optimal antenna placement in this case can be obtained in closed-form. However, for $M \geq 2$, the optimization problem becomes significantly more complex for the following two main reasons. First, the channel vectors are intricately dependent on the pinching-antenna positions, which themselves are variables to be optimized. Second, the objective function involves a non-trivial coupling between the pinching-antenna positions and the beamforming vectors.

To address the complexity of problem \eqref{p: sum rate maximization}, a weighted minimum mean square error (WMMSE)-based approach was introduced in \cite{xu2025pinching}. By leveraging the WMMSE framework, the original sum-rate maximization problem can be reformulated as follows:
\begin{subequations} \label{p: wmmse problem}
    \begin{align}
        \underset{\tilde \xb, \wb,\vb_m, \forall m}{\rm minimize} ~&\sum_{m=1}^M (w_m e_m - \log w_m) \\
        \st~ & \sum_{m=1}^M \|\mathbf{v}_{m}\|^2 \leq P_{\max}, \\
        & 0 \leq \tilde x_n \leq x_{\max}, \ \ \forall n \in \Nset,
    \end{align}
\end{subequations}
where $e_m$ denotes the mean square error of estimating the information of user $m$, while $\wb = [w_1,\ldots,w_M]^\top$ are the minimum mean square error (MMSE) weight.
Compared to problem \eqref{p: sum rate maximization}, the WMMSE reformulation offers a more tractable structure: the objective function is convex with respect to the beamforming vectors $\vb_m$ and the MMSE weights, though not necessarily in the antenna positions $\tilde x_n$. In contrast, the original problem contains fractional forms involving the channel vectors and beamformers, leading to a highly coupled and challenging optimization problem.

Because the optimization variables are decoupled in the constraints, the WMMSE problem can be iteratively solved by the BCD approach. Nevertheless, as highlighted in \cite{xu2025pinching}, optimizing the pinching-antenna positions remains non-convex, and a linear search is typically needed, which can result in substantial computational overhead for systems with long waveguides or high-frequency operation. To mitigate this issue, a low-complexity two-stage algorithm was proposed in \cite{xu2025pinching}. In the first stage, an MRT-based method is used to efficiently determine the pinching antenna positions. With these locations fixed, the second stage applies the standard WMMSE algorithm to optimize the beamforming vector. This two-stage strategy circumvents the need to perform a linear search over the antenna positions in each iteration of the WMMSE algorithm, thereby greatly reducing the overall computational complexity while maintaining near-optimal performance.

Next, we evaluate the achievable sum data rate of pinching-antenna systems with multiple waveguides, using a conventional fixed-position antenna system as a benchmark. For a fair comparison, the reference system employs a uniform linear array (ULA) at the base station with the same number of antennas ($M$) as the pinching-antenna system. The ULA is centered at the same location as the feed point, $[0,0,d_v]$, and uses a standard half-wavelength antenna spacing, $\Delta = \frac{\lambda}{2}$. Since the positions of the reference antennas are fixed, only the digital beamformers are optimized, and the sum-rate maximization is performed using the standard WMMSE algorithm.

\begin{figure}[!t]
	\centering
	\includegraphics[width=0.92\linewidth]{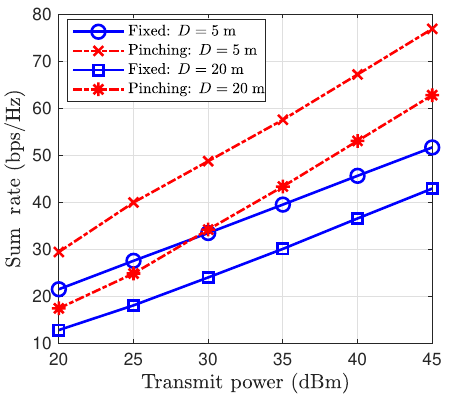}\\
        \captionsetup{justification=justified, singlelinecheck=false, font=small}	
        \caption{Achievable sum rate versus total transmit power $P_{\max}$ with $M=P=8$.} \label{fig: fix rate power} 
\end{figure}

Fig. \ref{fig: fix rate power} compares the achievable sum rates of the pinching-antenna system and the fixed-position antenna system for two representative coverage areas, $D \in \{5, 20\}$ m, across a range of transmit power levels. The results show that the pinching-antenna system consistently outperforms the fixed-position antenna system for all considered power levels. This substantial performance gain is attributed to the ability of pinching antennas to strengthen LoS links while simultaneously reducing inter-user interference.

In addition to classical optimization approaches, recent work has explored machine learning-based methods for joint beamforming and pinching-antenna position optimization in multi-waveguide, multi-user systems. For example, the authors of \cite{zhou2025gradient} considered a generalized weighted sum-rate maximization problem which is given by
\begin{subequations} \label{p: wsr meta}
    \begin{align}
        \underset{\tilde \xb, \{\vb_m\}}{\maxn}~ &\sum_{m=1}^M \omega_m \log_2\left(1 + \frac{|\hb_m^\top \vb_m|^2}{\sum_{i \neq m}|\hb_m^\top \vb_i|^2 + \sigma_m^2}\right) \\
        \st~ & \sum_{m=1}^M \|\vb_m\|^2 \leq P_{\max}, \\
        & \log_2\left(1 + \frac{|\hb_m^\top \vb_m|^2}{\sum_{i \neq m}|\hb_m^\top \vb_i|^2 + \sigma_m^2}\right) \geq R_m^{\min}, ~\forall m, \\
        & 0 \leq \tilde x_n \leq x_{\max},~ \forall n \in \Nset,
    \end{align}
\end{subequations}
where $\omega_m$ denotes a user-specific weight and $R_m^{\min}$ is the minimum required data rate for user $m$.

This setting is more challenging than the standard sum-rate maximization problem, due to the introduction of additional QoS requirements and more complex coupling between variables. Traditional BCD-type methods can become computationally prohibitive as network size and constraint set grow. To tackle this challenge, a gradient-based meta-learning joint optimization algorithm was proposed in \cite{zhou2025gradient}. The core idea is to ``learn to optimize'': two neural networks are trained, one to update the beamforming vectors ${\vb_m}$ and another one to update the pinching-antenna positions $\tilde{\xb}$, by mimicking a gradient-based optimization process across a diverse set of channel realizations and QoS scenarios. During training, the networks are optimized to minimize the average negative weighted sum-rate, with the power and minimum rate constraints incorporated via penalty terms in the loss function. Once trained, the meta-learner can rapidly produce near-optimal beamforming and antenna placement solutions for new, unseen system configurations, yielding significant improvements in computational efficiency and scalability. Numerical results demonstrate that the meta-learning approach achieves better weighted sum-rate performance and rapid convergence compared to classical methods, particularly in scenarios with stringent QoS requirements. These findings indicate that data-driven and meta-learning-based algorithms are promising to tackle complex, large-scale joint optimization problems for pinching-antenna systems.

\subsubsection{Multiple Pinching Antennas Per Waveguide}
This subsection focuses on a more general scenario where multiple pinching antennas are deployed along each waveguide, as shown in Fig. \ref{fig: system model multiple waveguide}(b). Without loss of generality, assume that each waveguide is equipped with $N$ pinching antennas. 
Denote the position of pinching antenna $n$ on waveguide $p$ by $\widetilde \psib_{p,n}^{\pin} = [\tilde x_{p,n}, 0, d_v]$.
Analogous to \eqref{eqn: channel multiple pinching antenna}, the channel between the pinching antennas on waveguide $p$ and user $m$ is given by
\begin{align} \label{eqn: channel multiple waveguide 2}
    h_{p,m} = \sum_{n=1}^N \frac{\eta^{\tfrac{1}{2}} e^{-j\big(\frac{2 \pi}{\lambda}[(\tilde x_{p,n} - \bar x)^2 + C_{m,p} ] + \frac{2 \pi}{\lambda_g}\tilde x_{p,n}\big)}}{\big[(\tilde x_{p,n} - \bar x)^2 + C_{m,p} \big]^{-\frac{1}{2}}},
\end{align}
for $p \in \mathcal{P}, m \in \Mset$. 
Comparing \eqref{eqn: channel multiple waveguide 2} with \eqref{eqn: channel multiple waveguide 1}, one can see that deploying more antennas on one waveguide introduces additional degrees of freedom for adjusting the channel.

Based on the channel model in \eqref{eqn: channel multiple waveguide 2}, several recent studies have investigated pinching-antenna-enabled MIMO communications \cite{wang2025modeling,zhao2025waveguide,guo2025gpass,xu2025joint,xu2025joint1,zhang2025two,xu2025pinching1,shan2025exploiting,sun2025multiuser,wang2025sum,shan2025multigroup,zhao2025resource}.
Specifically, the authors of \cite{wang2025modeling} investigated energy-efficient pinching-antenna system designs by jointly optimizing the pinching-antenna positions and downlink beamformers. To solve the challenging optimization problem, an iterative algorithm was proposed based the BCD and SCA methods. The study demonstrates substantial transmit power savings compared to conventional massive MIMO systems. A staged GNN architecture has been proposed in \cite{guo2025gpass}, which efficiently learns both antenna placement and beamforming, enabling scalable and accurate resource allocation for pinching-antenna systems with strong generalization across diverse network sizes and user distributions. Meanwhile, the authors of \cite{xu2025joint} introduced a Karush-Kuhn-Tucker (KKT)-guided dual learning framework based on a transformer architecture, enabling rapid and high-quality joint optimization of transmit and pinching beamforming in multi-user pinching-antenna systems.
The authors of \cite{bereyhi2025mimo} explored the use of pinching-antenna systems for uplink and downlink multi-user MIMO communication. The digital precoder and pinching-antenna positions are jointly optimized to maximize the weighted sum-rate. Two low-complexity iterative algorithms are proposed based on fractional programming (FP) and Gauss-Seidel methods. The results show that pinching-antenna systems offer significant throughput gains over conventional massive MIMO and hybrid analog-digital systems, highlighting their promise for next-generation wireless networks.
A recent study \cite{wang2025antenna} investigated a practical implementation of pinching-antenna systems in multi-waveguide scenarios, where antennas are installed at pre-configured discrete locations rather than being continuously adjustable. Focusing on downlink NOMA transmission, the authors formulated a joint optimization problem involving waveguide assignment, antenna activation, SIC decoding order, and power allocation. By modeling assignment and activation as coalition-formation games and applying monotonic optimization and SCA, both globally optimal and near-optimal low-complexity solutions are obtained.

Recently, the pinching antenna-enabled multicast transmission was investigated in \cite{chen2025dynamic}, where the dynamic pinching antenna placement was studied in a downlink multi-user multiple-input-single-output (MU-MISO) setting. A cross-entropy optimization framework was proposed for both single-group and multi-group multicast scenarios, aiming to maximize the minimum data rate of the users. It is observed that flexible antenna positioning of pinching-antenna system can  improve multicast beamforming efficiency, particularly under varying user locations. Simulation results verify that dynamic positioning significantly outperforms static schemes, especially when the number of users increases.

Most existing studies on pinching-antenna systems have focused on single-cell setups, overlooking the inter-cell interference arising from spatially distributed waveguides connected to different BSs. To address this limitation, the authors of \cite{sun2025stochastic} developed an analytical framework for multi-cell pinching-antenna systems based on stochastic geometry. By modeling the spatial distribution of BSs and users, this study derived closed-form expressions for the outage probability and validated the analysis through simulations. The results reveal that, even under multi-cell interference, pinching-antenna systems significantly outperform conventional fixed-antenna counterparts, thereby demonstrating their strong potential in large-scale network deployments.

The main design principles and takeaways for multiple-waveguide pinching-antenna systems are summarized below:

\begin{tcolorbox}
\begin{itemize}\itemsep = 1mm
\item \textbf{Enhanced Spatial Multiplexing:} Leveraging multiple waveguides provides greater degrees of freedom for pinching antenna deployment, substantially boosting spatial diversity, user capacity, and spectral efficiency compared to single-waveguide or fixed-antenna systems.
\item \textbf{System Design Challenges and Solutions:} Achieving optimal performance requires joint optimization of pinching-antenna positions and beamforming, resulting in highly coupled and non-convex problems. Iterative algorithms, such as BCD and WMMSE/FP-based methods, are effective in tackling these challenges. Nevertheless, developing more practical and scalable algorithms for real-time and large-scale deployment remains an important research direction.
\item \textbf{Pinching Antenna Activation:} In addition to pinching-antenna optimization, efficient antenna activation, i.e., activating pinching antennas at pre-configured positions is a practical and important research direction, which enables flexible system deployment while reducing implementation complexity.
\end{itemize}
\end{tcolorbox}

Table \ref{tab:pass-design-summary} provides a structured summary of representative system design problems and corresponding design strategies for pinching-antenna systems. It highlights the diversity of architectures, optimization variables, and transmission schemes addressed in the literature. This comparative overview underscores the adaptability of the pinching-antenna framework, the progression from closed-form designs to advanced iterative and learning-based algorithms, and the importance of system complexity and application context in shaping the choice of solution strategy. These foundations set the stage for addressing more challenging and practical issues in future research on generalized pinching-antenna systems.


\begin{table*}[ht]
\begin{threeparttable}
\caption{Summary of representative system design problems and solutions for pinching-antenna systems.}
\centering
\renewcommand{\arraystretch}{1.7}
\begin{tabular}{|
    >{\centering\arraybackslash}p{1.5cm}|
    >{\centering\arraybackslash}p{3cm}|
    >{\centering\arraybackslash}p{1.6cm}|
    >{\centering\arraybackslash}p{1.6cm}|
    >{\centering\arraybackslash}p{3.1cm}|
    >{\centering\arraybackslash}p{4.1cm}|
    }
\hline
\multicolumn{3}{|c|}{\textbf{System setup}} & \multicolumn{3}{c|}{\textbf{Problem \& Solutions}} \\
\hline

\textbf{Number of Waveguides}
    & \textbf{Number of pinching antennas per waveguide}
    & \textbf{Number of users}
    & \textbf{Transmission scheme}
    & \multirow{2}{*}{\textbf{Variables to be optimized}}
    & \multirow{2}{*}{\textbf{Viable solutions}} \\
\hline

\multirow{9}{*}{Single}
  & \multirow{4}{*}{Single}
     & Single   & -  & Antenna position  & Closed-form \cite{xu2025pinching,ding2025los,xu2025pinching_los,tyrovolas2025performance} \\
  \cline{3-6}
  &                                             & \multirow{3}{*}{Multiple} & \multirow{2}{*}{OMA} & Antenna position, power allocation & \multirow{2}{*}{Closed-form \cite{ding2025analytical}}\\
  \cline{4-6}
  & & & \multirow{2}{*}{NOMA}  & Antenna position, power allocation & Closed-form \cite{ding2025analytical,xu2025qos}; BCD + SCA \cite{zeng2025energy1} \\
  \cline{2-6}
  & \multirow{5}{*}{Multiple}
     & Single  & - & Antenna position   & Closed-form \cite{xu2025rate,xie2025low} \\
  \cline{3-6}
  &                                             & \multirow{3}{*}{Multiple} & \multirow{2}{*}{OMA}  & Antenna position, time allocation  & {BCD + SCA \cite{tegos2024minimum,zeng2025energy}; deep learning \cite{xie2025graph,karagiannidis2025deep}} \\
  \cline{4-6}
  &                                             & & \multirow{2}{*}{NOMA}  & Antenna position, power allocation & \multirow{2}{*}{BCD + SCA \cite{xu2025qos,hu2025sum,zeng2025energy1}} \\
\hline

\multirow{5}{*}{Multiple}
  & \multirow{3}{*}{Single}
     & Single   & -  & Antenna position  & Closed-form \cite{xu2025pinching} \\
  \cline{3-6}
  &                                             & \multirow{2}{*}{Multiple} & \multirow{2}{*}{SDMA}  & Antenna position, beamforming                    & {WMMSE/FP + BCD \cite{xu2025pinching,bereyhi2025downlink}; meta learning \cite{zhou2025gradient}} \\
  \cline{2-6}
  & \multirow{3}{*}{Multiple}
     & Single   & -  & Antenna position                         & - \\
  \cline{3-6}
  &                                             & \multirow{2}{*}{Multiple} & \multirow{2}{*}{SDMA}  & Antenna position, beamforming                    & {Penalty method + SCA \cite{wang2025modeling,zhao2025waveguide}; deep learning \cite{guo2025gpass,xu2025joint,he2025ris}} \\
\hline
\end{tabular} \label{tab:pass-design-summary}
\end{threeparttable}
\end{table*}

\subsection{Pinching-Antenna System Design for Random LoS and NLoS Channels} \label{sec: los_nlos}

In the previous subsections, we considered downlink pinching-antenna system design assuming the availability of an LoS channel. However, in more complex wireless environments, the LoS channel can be intermittently blocked by obstacles, and the NLoS component can dominate due to multipath scattering. Therefore, it is important to study the performance of pinching-antenna systems for random LoS blockage and NLoS fading, as these factors significantly affect system performance, especially in dynamic environments such as urban canyons or indoor areas with heavy obstruction. In this subsection, we investigate pinching-antenna system design for random LoS and NLoS channels.

\subsubsection{System Model and Channel Setup}

Consider a basic downlink system where a single pinching antenna serves multiple users using TDMA. The users are randomly deployed in a rectangular area of size $D_x \times D_y$. In TDMA, users are allocated orthogonal time slots, ensuring that interference is minimized. The position of the pinching antenna is optimized once for the whole scheduling duration for all users.
Without loss of generality, we consider a communication region where $M$ users are randomly distributed.

The channel between the pinching antenna and user $m$ is modeled as a combination of LoS and NLoS components. The total channel gain is given by 
\begin{align}
    h_m = \xi_m h_m^{\mathrm{LoS}} + h_m^{\mathrm{NLoS}},
\end{align}
where $\xi_m \in \{0, 1\}$ is a Bernoulli random variable indicating the presence ($\xi_m = 1$) or absence ($\xi_m = 0$) of a direct LoS path, $h_m^{\mathrm{LoS}}$ is the LoS channel coefficient defined in \eqref{eqn: channel model siso}, and $h_m^{\mathrm{NLoS}} \sim \mathcal{CN} \left(0, \frac{\mu_{m}^2}{r_m^2(\tilde x)}\right)$ is the NLoS channel component, modeled using a stochastic fading model for scattered multipath signals \cite{xu2025pinching_nlos}. Here, $\frac{\mu_{m}^2}{r_m^2(\tilde x)}$ denotes the distance-dependent NLoS channel power with $r_m(\tilde x)$ denoting the distance between user $m$ and the pinching antenna. The random variable $\xi_m$ follows a probability distribution that depends on the distance between the pinching antenna and the user, as well as environmental factors that affect the likelihood of blockage. Given the channel model above, the instantaneous received SNR at user $m$ is given by $\Gamma_m(\tilde x) = \rho_m |h_m(\tilde x)|^2$, where $\rho_m = \frac{P_{\max}}{\sigma_m^2}$ with $P_{\max}$ denoting the transmit power of the pinching antenna and $\sigma_m^2$ representing the received noise power at user $m$. The probability to maintain an LoS link between user $m$ and the pinching antenna is modeled as $\mathbb{P}(\xi_m=1) = e^{-\beta r_m^2(\tilde x)}$, where $\beta$ is an environment-dependent parameter reflecting the density of blockages.

\subsubsection{Optimization Problems}

To optimize the performance of the pinching-antenna system under random LoS and NLoS channels, we formulate two complementary optimization problems based on different performance metrics:

\textbf{2.1 Average SNR-based Design:}  
The average-SNR-based design focuses on maximizing the long-term average received SNR of the system by optimizing the antenna placement. We use the average SNR as a tractable surrogate for long-term throughput, since for a fixed modulation and coding scheme, the achievable rate is a monotonically increasing function of the received SNR, and averaging the SNR yields much simpler expressions than averaging $\log(1+\mathrm{SNR})$, which facilitates analysis and optimization. The goal is to maximize the minimum average SNR across all users, so that the system can simultaneously achieve good spectral-efficiency performance and preserve user fairness. The corresponding optimization problem is formulated as

\begin{subequations} \label{eq:avg_snr_opt}
\begin{align}
    \underset{\tilde x}{\maxn} \quad & \min_{1 \le m \le M} \bar \Gamma_m(\tilde x) \\
    \st \quad & 0 \le \tilde x \le D_x,
\end{align}
\end{subequations}
where $\bar \Gamma_m(\tilde x)$ is the average SNR of user $m$.

\textbf{2.2 Outage-Constrained Design:}  
The outage-constrained design aims to guarantee a minimum level of reliability for each user by constraining the outage probability. Instead of maximizing the instantaneous rate directly, we optimize an SNR threshold that serves as a design target for selecting modulation and coding schemes: a higher threshold allows more aggressive coding and higher rates, while still respecting reliability requirements. The goal is therefore to maximize this SNR threshold while ensuring the probability that the instantaneous SNR falls below the threshold is less than a prescribed value for each user. The outage-constrained problem is formulated as
\begin{subequations} \label{eq:outage_opt}
\begin{align}
    \underset{\tilde x, t}{\maxn} \quad & t \\
    \st \quad & \mathbb{P}\left(\Gamma_m(\tilde x) \ge t \right) \ge 1 - \varepsilon_m, \quad \forall m \in \mathcal{M} \\
    & 0 \le \tilde x \le D_x,
\end{align}
\end{subequations}
where $t$ is the target SNR threshold, and $\varepsilon_m$ is the prescribed outage probability for user $m$.

These two optimization problems capture the trade-off between maximizing throughput (average SNR) and ensuring reliability (outage-constrained SNR) in a stochastic LoS/NLoS environment, where random blockage and fading introduce additional challenges to system design. As shown in \cite{xu2025pinching_nlos}, problems \eqref{eq:avg_snr_opt} and \eqref{eq:outage_opt} can be efficiently solved by using bisection-based algorithms.

\subsubsection{Simulation Results}

\begin{figure}[!t]
	\centering
	\includegraphics[width=0.88\linewidth]{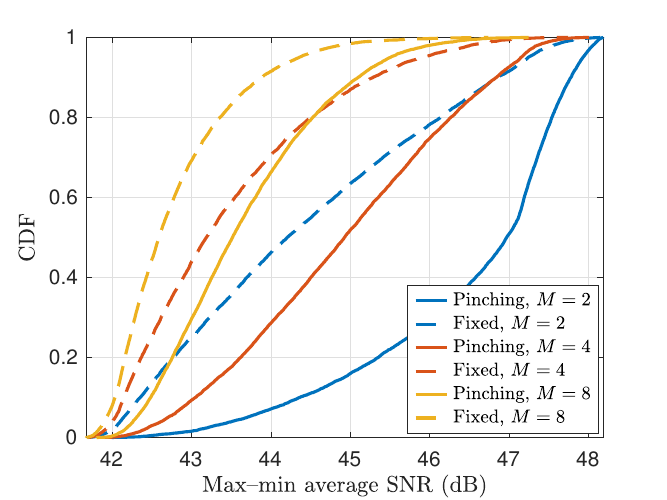}\\
        \captionsetup{justification=justified, singlelinecheck=false, font=small}	
        \caption{CDF of the max-min average SNR for pinching- and fixed-antenna systems with $D_x = 40$ m.}
    \label{fig:CDF_SNR}
\end{figure} 

\begin{figure}[!t]
	\centering
	\includegraphics[width=0.88\linewidth]{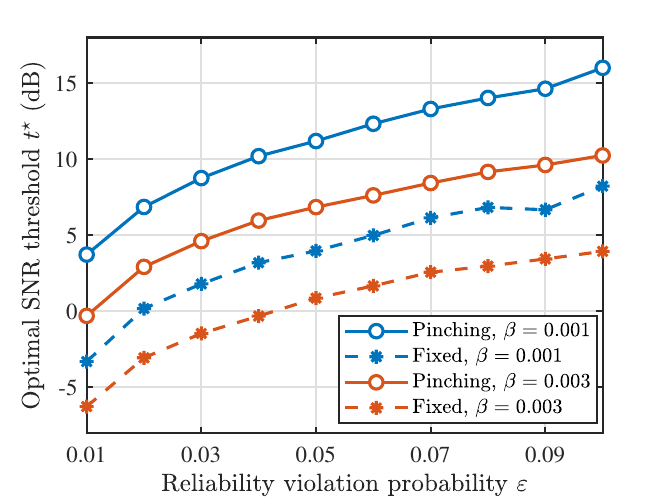}\\
        \captionsetup{justification=justified, singlelinecheck=false, font=small}	
        \caption{Optimal SNR threshold $t^*$ versus outage probability $\varepsilon_m = \varepsilon, \forall m,$ for pinching and fixed antenna systems with $D_x = 40$ m.}\vspace{-0mm} \label{fig:oc t epsilon} 
\end{figure} 

To demonstrate the performance of the pinching-antenna system under random LoS and NLoS conditions with the max-min average SNR metric, we simulate the cumulative distribution functions (CDFs) of the max-min average SNR for both pinching-antenna and fixed-antenna systems, as shown in Fig.~\ref{fig:CDF_SNR}. In the simulation, we consider different numbers of users, $M \in \{2,4,8\}$ and we set $\mu_m^2 = -80$ dBm, $\sigma_m^2 = -90$ dBm, communication area side length of $D_y = 10$ m and $\beta = 0.001$. As observed, the proposed pinching-antenna system consistently achieves higher SNR across all percentiles compared to its fixed counterpart, indicating a clear improvement in both typical and worst-case performance.

To evaluate the performance of the pinching-antenna system under the outage-constrained metric, we plot the optimal SNR thresholds $t^*$ for both pinching- and fixed-antenna systems as a function of outage probability, considering two different LoS blockage coefficients, $\beta$ in Fig. \ref{fig:oc t epsilon}. A larger $\beta$ indicates a larger LoS blockage probability. As expected, $t^*$ increases monotonically with the outage probability $\varepsilon$ for both systems, as a more relaxed reliability constraint allows higher transmission thresholds. At a given reliability level, the pinching-antenna system consistently outperforms the fixed-antenna system, demonstrating its ability to adaptively reposition and improve the LoS condition for users. 

Overall, these results provide quantitative evidence that pinching antennas can significantly enhance link robustness and reliability, even in challenging blockage-dominated environments.

\subsection{Uplink Pinching-Antenna Systems}
While the preceding subsections primarily focus on downlink pinching-antenna system design, the uplink scenario, where users transmit toward a central access point via pinching antennas, has received comparatively limited attention. Nevertheless, uplink design is both practically important and theoretically distinct, due to unique resource allocation, signal detection, and interference management challenges, especially when multiple pinching antennas are deployed along the same waveguide.

Recent literature has begun to address these challenges from several perspectives \cite{tegos2024minimum,zeng2025energy1,bereyhi2025mimo,ouyang2025uplink2}. The first detailed study of the multi-user uplink in pinching-antenna systems was in \cite{tegos2024minimum}, where TDMA is used to avoid inter-channel interference. The authors formulated and solved a joint optimization problem to determine the radio resources allocation and pinching-antenna positions  to maximize the minimum data rate among users, thereby ensuring fairness and meeting diverse QoS requirements. To further improve spectral efficiency, NOMA-based uplink pinching-antenna systems were studied in \cite{zeng2025energy1}.
Considering a MIMO setting, the authors of \cite{bereyhi2025mimo} investigated the joint design of both uplink and downlink transmission for pinching-antenna systems. The sum-rate maximization problem was formulated and solved using an iterative algorithm. Simulation results demonstrated the performance gain of pinching-antenna systems in uplink MIMO scenarios as compared to conventional fixed-antenna systems.

However, earlier works implicitly assumed that uplink signals could be perfectly delivered from each pinching antenna to the central processor, overlooking a critical challenge in multi-antenna deployments. Specifically, when multiple pinching antennas are deployed along a single dielectric waveguide in the uplink, inter-antenna radiation leakage can occur: signals received by one pinching antenna may unintentionally propagate and be re-radiated through other antennas as they travel toward the feed point. This phenomenon introduces unintended interference, complicates analytical modeling, and can degrade overall system performance.

To address this issue, a recent study in \cite{ouyang2025uplink2} proposed a segmented waveguide-enabled pinching-antenna architecture. In this design, the traditional long dielectric waveguide is divided into multiple shorter segments, each equipped with its own feed point and associated pinching antennas. By ensuring that only one pinching antenna is active per segment, this approach effectively eliminates inter-antenna radiation leakage and enables a tractable, physically consistent uplink signal model that supports multi-antenna operation. The work further introduced three practical operating protocols, including the segment selection, the segment aggregation, and the segment multiplexing, each offering a distinct trade-off in terms of performance and implementation complexity.

\begin{figure}[!t]
	\centering
	\includegraphics[width=0.92\linewidth]{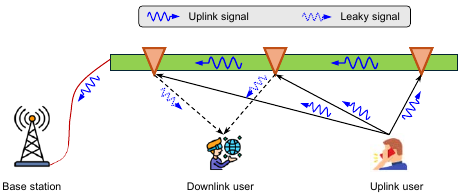}\\
        \captionsetup{justification=justified, singlelinecheck=false, font=small}	
        \caption{Illustration of an uplink pinching-antenna system with multiple antennas and the signal leakage issue during transmission.} \label{fig: system model isac} 
\end{figure} 

From a practical standpoint, waveguide segmentation also introduces non-negligible implementation implications. A continuous dielectric waveguide represents the lowest-complexity baseline, where a single low-loss dielectric waveguide is deployed along the service area and pinching antennas are realized as local radiating points without additional active components. By contrast, segmented-waveguide solutions require extra RF components, such as connectors, switches, or reconfigurable couplers, to isolate or selectively feed different segments. While this segmentation helps mitigate uplink inter-antenna radiation leakage, it inevitably increases hardware and installation complexity. Moreover, segmentation introduces an additional algorithmic design dimension: The system must decide which waveguide or segment to activate based on the spatial distribution and traffic demands of the users, motivating further research on user-aware waveguide selection and scheduling strategies in practical deployments.

\subsection{Control and Implementation of Pinching-Antenna Positioning}

In the previous subsections, we mainly focused on where the pinching antenna should be placed by solving optimization problems. In practice, these designs must be realized through a control mechanism that adjusts the pinching-antenna position along the waveguide. Depending on the implementation, there are two relevant approaches for implementing pinching-antenna control: continuous-position and discrete-position.

In a continuous-position implementation, the pinching antenna is mounted on a rail or slider, enabling its position along the waveguide to be continuously adjusted. This supports both a real-time user-centric mode of operation and a long-term environment-aware mode of operation. In the real-time mode, the antenna position is adapted according to the instantaneous or short-term estimated locations of the active users, which can be obtained using methods such as cellular positioning, device-based sensing, or machine learning-based localization techniques. In the long-term mode, the antenna position is optimized based on the environment and network-level performance, for example by accounting for the spatial user distribution, blockage patterns, and QoS requirements in the service area. In this case, the antenna position is updated only occasionally when the environment changes, and serves as a slowly varying configuration knob that complements conventional beamforming and resource allocation.

For a discrete-position implementation, multiple pinching antennas are pre-deployed at specific locations along the waveguide. The control task is to decide which antenna(s) to activate based on the real-time user distribution (real-time mode) or based on static environmental factors such as traffic patterns and persistent blockage (long-term mode). This activation-based control can be implemented electronically and supports moderate to fast reconfiguration.

The methods presented in this work primarily correspond to the real-time mode of operation for both continuous and discrete positioning, where antenna positions (or active antennas) are updated dynamically based on user locations or large-scale channel information. In practice, such real-time operation can be realized by segment switching in discrete-position implementations and by small mechanical motions in continuous-position implementations, scheduled over a finite set of candidate activation points and operating on the time scales of typical human or vehicular mobility rather than at the symbol level. At the same time, the same optimization framework can be applied in an environment-aware manner by replacing instantaneous user locations with statistical information (e.g., traffic and blockage patterns), which is particularly attractive for outdoor or highly dynamic scenarios where per-user tracking is difficult and pinching antennas are used to enhance coverage and throughput at the network level rather than to follow individual users.


\section{Broader Horizons: Versatile and Emerging Applications of Pinching Antennas} \label{sec: application}
The previous sections have focused on the core system architectures and design methodologies of pinching-antenna systems; however, we note that the potential of this flexible-antenna technology extends far beyond conventional communication scenarios. Recent advances have demonstrated that pinching antennas can be seamlessly integrated with other cutting-edge wireless techniques, enabling new capabilities and opening up a range of emerging applications. In this section, we explore several representative directions, including ISAC, cooperative transmission frameworks, and synergistic designs with advanced wireless technologies, that highlight the versatility and transformative promise of pinching-antenna systems for next-generation networks.

\subsection{Enabling ISAC with Pinching Antennas}

ISAC has emerged as a transformative technology for next-generation wireless networks, enabling seamless integration of communication and sensing functionalities in a unified infrastructure \cite{liu2022integrated,zhang2021overview,zhang2024optimal,ma2025deep,wang2024integrated,wang2025cooperative}. By jointly utilizing spectrum, hardware, and signal processing resources, ISAC is expected to support a wide range of applications, such as high-precision localization, environment monitoring, and smart industrial automation. Recent research has identified ISAC as a key use case for 6G and beyond, attracting significant attention from both academia and industry \cite{you2021towards}.

Pinching-antenna systems offer unique advantages that make them particularly promising for ISAC applications. By flexibly activating and positioning antennas along dielectric waveguides, pinching-antenna systems can establish robust LoS links, reduce large-scale path loss, and dynamically adapt the array configuration to meet diverse service requirements \cite{ding2025pinching,mao2025multi,qin2025joint,zhang2025integrated,khalili2025pinching,wang2025wireless,ouyang2025rate,bozanis2025cram,illi2025secure,li2025pin-isac}. The inherent reconfigurability and low-cost implementation of pinching antennas also enable scalable and user-centric deployment, which is highly desirable for ISAC tasks. As recent studies have demonstrated, these capabilities not only enhance communication rates, but also significantly improve sensing performance, such as positioning accuracy and target detection, compared to conventional fixed-antenna systems. Thus, pinching-antenna systems present a highly versatile technique for advancing ISAC in future wireless networks. In what follows, we will demonstrate how to use pinching antennas to facilitate ISAC from the system design perspective.

Consider a pinching-antenna-enabled ISAC system where $P$ transmit pinching antennas are jointly used to serve a single-antenna user and $Q$ receive pinching antennas are exploited to sensing a potential point target using the received signal reflected from the target, as shown in Fig. \ref{fig: system model isac}. Without loss of generality, we assume that each of the pinching antennas is deployed on a different waveguide. The positions of the $p$-th transmit pinching antenna and the $q$-th receive pinching antenna are denoted by $\psib^{\pin}_{T,p} = [\tilde x_{T,p}, \tilde y_{T,p}, d_v]$ and $\psib^{\pin}_{R,q} = [\tilde x_{R,q}, \tilde y_{R,q}, d_v]$, respectively. The positions of the user and the target are represented by $\psib_u = [x_u,y_u,0]$ and $\psib_t = [x_t,y_t,0]$, respectively. Then, the channel between the $p$-th transmit pinching antenna and the target and the channel between the target and the $q$-th receive pinching antenna are respectively represented by
\begin{subequations}
    \begin{align}
        h_{t,p}(\tilde x_{T,p}) &= \frac{\eta^{\frac{1}{2}} e^{-j\left(\frac{2\pi}{\lambda}\|\psib^{\pin}_{T,p} - \psib_t\| + \frac{2\pi}{\lambda_g}\tilde x_{T,p} \right)} }{\|\psib^{\pin}_{T,p} - \psib_t\|}, \\
        g_{t,q}(\tilde x_{R,q}) &= \frac{\eta^{\frac{1}{2}} e^{-j\left(\frac{2\pi}{\lambda}\|\psib^{\pin}_{R,q} - \psib_t\| + \frac{2\pi}{\lambda_g}\tilde x_{R,q} \right)} }{\|\psib^{\pin}_{R,q} - \psib_t\|},
    \end{align}
\end{subequations}
where we assume that the feed points of the $p$-th transmit pinching antenna and the $q$-th receive pinching antenna are $\widehat \psib^{\pin}_{T,p} = [0, \tilde y_{T,p}, d_v]$ and $\widehat\psib^{\pin}_{R,q} = [0, \tilde y_{R,q}, d_v]$, respectively. Based on the above channel model, the received sensing signals at the $Q$ receive pinching antennas can be represented by
\begin{align}
    \yb_s = \zeta \gb_t(\{\tilde x_{R,q}\}) \hb_t^{\Hf}(\{\tilde x_{T,p}\}) \wb s + \nb_s,
\end{align}
where $\zeta$ denotes the reflection coefficient of the target, $\wb$ denotes the transmit beamformer of the $P$ transmit pinching antennas, $\gb_t(\{\tilde x_{R,q}\}) = [g_{t,1}(\tilde x_{R,1}),...,g_{t,Q}(\tilde x_{R,Q})]$, and $\hb_t(\{\tilde x_{T,p}\}) = [h_{t,1}(\tilde x_{T,1}),...,h_{t,P}(\tilde x_{T,P})]$. Furthermore, $s \sim \CN(0,1)$ and $\nb_s \sim \CN({\bf{0}}, \sigma_s^2 \Ib)$ denote the transmit signal and the received noise at the $Q$ receive pinching antennas, respectively, where $\sigma_s^2$ represents the noise power. 

\begin{figure}[!t]
	\centering
	\includegraphics[width=0.92\linewidth]{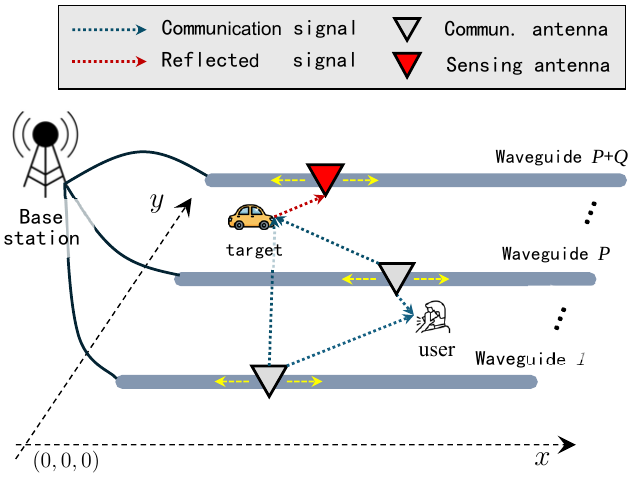}\\
        \captionsetup{justification=justified, singlelinecheck=false, font=small}	
        \caption{Pinching-antenna-enabled ISAC system.} \label{fig: system model isac} 
\end{figure} 

As noted in \cite{mao2025multi}, to enhance the received signal strength reflected from the target, the receive pinching antennas should be placed on the waveguide closest to the target, i.e., $\tilde x_{R,q} = x_t, \forall q$. By applying a receive beamfomer, $\vb = \frac{\gb_t(\{\tilde x_{R,q}\})}{\|\gb_t(\{\tilde x_{R,q}\})\|}$, at the $Q$ receive pinching antennas, the achievable sensing SNR can be written as follows
\begin{align}
    \Gamma_s(\wb,\{\tilde x_{T,p}\}) = \left(\sum_{q=1}^Q \frac{\eta \zeta^2}{(\tilde y_{R,q} - y_t)^2 + d_v^2}\right) \frac{\|\hb_t^{\Hf}(\{\tilde x_{T,p}\}) \wb\|^2}{Q \sigma_s^2}.
\end{align}
Based on the above model, the problem to maximize the communication data rate subject to the sensing SNR and transmission power constraints can be formulated as
\begin{subequations} \label{eqn: isac}
    \begin{align}
        \underset{\wb, \{\tilde x_{T,p}\}}{\maxn} ~& \log (1+\Gamma_c(\wb, \{\tilde x_{T,p}\})) \\
        \st~& \Gamma_s(\wb,\{\tilde x_{T,p}\}) \geq \bar \Gamma_s, \\
        & \wb^\Hf \wb \leq P_{\max},\\
        & 0 \leq \tilde x_{T,p} \leq \tilde x_{\max}, \, \forall p,
    \end{align}
\end{subequations}
where $\Gamma_c(\wb, \{\tilde x_{T,p}\})$ denotes the communication SNR, $\bar \Gamma_s$ denotes the minimum sensing SNR requirement, and $P_{\max}$ is the maximum transmission power.

\begin{figure}[!t]
	\centering
	\includegraphics[width=0.92\linewidth]{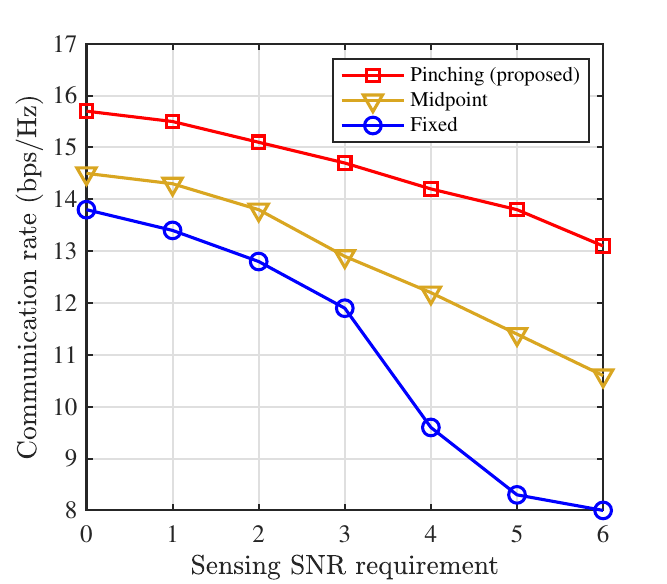}\\
        \captionsetup{justification=justified, singlelinecheck=false, font=small}	
        \caption{Achievable communication rate versus sensing SNR requirement in pinching-antenna-enabled ISAC system.} \label{fig: simulation isac} 
\end{figure}

Problem \eqref{eqn: isac} is highly non-convex and challenging to solve, as both the communication SNR and sensing SNR depend on the positions of the transmit pinching antennas in a complex manner. To address the joint optimization of the transmit beamformer and the antenna positions, the authors of \cite{mao2025multi} proposed an SCA-based algorithm to efficiently solve problem \eqref{eqn: isac}. The effectiveness of the considered pinching-antenna-enabled ISAC system is evaluated through numerical simulations, as shown in Fig. \ref{fig: simulation isac}. The results demonstrate that, for a given sensing SNR requirement, the pinching-antenna system significantly outperforms the fixed-antenna system in terms of achievable communication data rates. Furthermore, when compared to the ``Midpoint'' benchmark, where the transmit pinching antennas are simply placed at the midpoint between the user and the target along the waveguide ($\tilde x_{T,p} = \frac{x_u + x_t}{2}, \forall p$), the proposed SCA method achieves even greater performance, underscoring the importance of carefully optimizing the placement of the transmit pinching antennas.

Beyond the optimization of joint beamforming and antenna placement, several recent studies have provided additional insights for pinching-antenna-enabled ISAC systems. For example, the positioning capabilities of pinching antennas from the Cram\'{e}r–Rao lower bound (CRLB) perspective were studied in \cite{ding2025pinching}, where the analysis shows that such systems can notably reduce the CRLB of ISAC systems. The authors of \cite{qin2025joint} formulated a joint antenna positioning and power allocation problem for ISAC, proposing a maximum entropy-based reinforcement learning algorithm to achieve robust performance under stringent sensing and energy constraints. By introducing an outage-based reliability metric, the authors of \cite{khalili2025pinching} focused on improving the sensing performance by leveraging the spatial diversity of dynamically activated pinching antennas. The rate region of pinching-antenna-enabled ISAC systems was studied in \cite{ouyang2025rate}, where closed-form expressions for the achievable communication and sensing rates were derived. Collectively, these works highlight the diverse potential of pinching-antenna technology for advancing ISAC in terms of accuracy, reliability, flexibility, and performance trade-offs.

\subsection{Facilitating Cooperative Communications with Pinching Antennas}


With the proliferation of connectivity and new applications such as autonomous driving and smart homes, communication architectures are moving towards to decentralized with cooperative communications. Distributed antennas have emerged as a practical solution for improving coverage and capacity in future communication networks, by deploying spatially separated antennas connected to the BS  \cite{Saleh87distributed}.  To fully realize the potential of distributed antenna system (DAS), cooperative transmission between the BS and spatially distributed antennas is essential \cite{Heath13a}. By strategically coordinated signal processing, DAS can significantly enhance coverage, especially in environments with severe shadowing or penetration loss, such as dense urban areas and large indoor spaces \cite{DASWG23design,ma24tcom}. This makes DAS a compelling technology for next-generation wireless communications.

Pinching-antenna systems represent a natural evolution of traditional DAS, where pinching antennas are deployed along spatially separated waveguides that are connected to the BS via an interface unit. In contrast to conventional DAS deploying antennas using coaxial cable or fiber, pinching antennas provide a flexible and cost-effective solution by deploying low-complexity antennas along physical waveguides. Cooperative transmission between the BS and pinching antennas enables to collaboratively serve users by exploiting spatial diversity and distributed array gain. Through cooperative transmission, the BS and the pinching antennas can jointly transmit the signal using both NLoS and LoS paths, improving communication reliability and coverage robustness. Moreover, cooperative transmission enables constructive signal combining from the BS and pinching antennas by proper deployment of pinching antennas, yielding increased the received SINR or improving spatial multiplexing. Therefore, the cooperative transmission between the BS and the pinching antennas provides scalable and distributed communication, combining the strengths of centralized processing of the BS with the spatial agility of pinching antennas.

\begin{figure}[!t]
	\centering
	\includegraphics[width=0.92\linewidth]{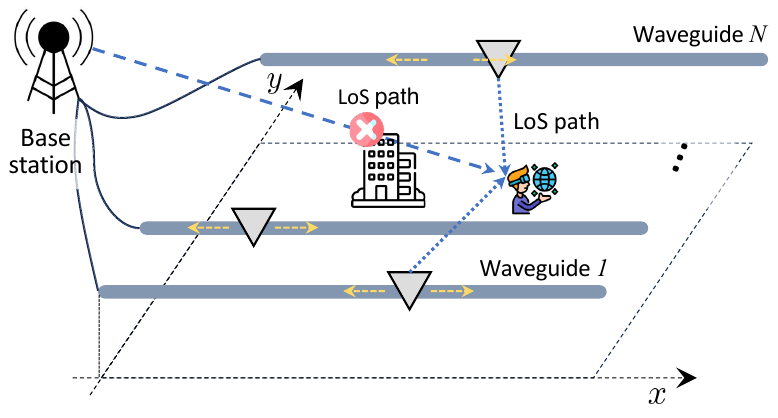}\\
        \captionsetup{justification=justified, singlelinecheck=false, font=small}	
        \caption{Pinching-antenna enabled cooperative system.} \label{fig:cooperative-pas} 
\end{figure}

Consider a cooperative transmission scenario  as shown in Fig. \ref{fig:cooperative-pas}, where a BS equipped with $N_B$ antennas communicates with a single-antenna user, assisted by a set of pinching antennas. Specifically, $P$ waveguides are deployed, each supporting $N$ active  pinching antennas. The LoS path between the BS and the user is blocked by buildings. Therefore, the user receives the signals from the BS via NLoS paths and from the pinching antennas via LoS paths.  Let the channel between the BS and the user be modeled as ${{\bm{h}}_B} = \sqrt \frac{{ \eta  }}{L_B^{\xi_B }}{\widetilde {\bm{h}}_B}$, where $L_B$ is the distance between the BS and the user, $\xi_B$ is the path loss coefficient, and $\widetilde {\bm{h}}_B$ denotes the Rayleigh fading component. 

Without loss of generality, the coordinate of the signal injection point for waveguide $p$ is denoted by $\bm{\psi}_{0,p} = [0,(p-1)d_h - \frac{D}{2},d_v], p\in \mathcal{P} \triangleq \{1,...,P\}$, for notation consistency in Section \ref{sec:Multiple-Waveguide Systems}. 
The coordinate of the pinching antenna on the $p$-th waveguide is denoted by $\widetilde\psib_{p,n}^{\pin} = [\tilde{x}_p, (p-1) d_h - \frac{D}{2}, d_v]$ and the position of the user is denoted by $\psib = [\bar x, \bar y,d_v]$. The phase delay experienced by the signal injected into waveguide $p$ when propagating through the $n$-th pinching antenna to reach the user can be expressed as 
\begin{equation}
{\psi_{p,n}} =  {\frac{{2\pi }}{\lambda }\left\| {\widetilde\psib_{p,n}^{\pin} -\psib} \right\| + \frac{{2\pi }}{{{\lambda_g}}}\left\| { \widetilde\psib_{p,n}^{\pin}- \bm{\psi}_{0,p} } \right\|}.
\end{equation}
Additionally, considering the analysis in \cite{xu2025pinching}, we neglect the in-waveguide attenuation.
Thus, the channel between the pinching antennas and the user through the $p$-th waveguide is given by 
\begin{equation}
\begin{aligned}
{h_{G,p}} &= 
\frac{{\sqrt \eta  }}{{\left\| \widetilde\psib_{p,n}^{\pin} -\psib \right\|}} \times \sqrt {\frac{1}{P}} \sum\limits_{n = 1}^{P} {{e^{ - j{\psi _{p,n}}}}},
\end{aligned}
\end{equation}
and we define $\mathbf{h}_{G} = [h_{G,1},\cdots,h_{G,P}]$ as the channel vector from the $P$ waveguides to the user. 
Then, the sum-rate maximization problem can be formulated as 
\begin{subequations} \label{eq:opt_cooperative_pinching}
    \begin{align}
    \underset{\mathbf{w},\{\tilde{x}_{p,n}\}}{\maxn} ~& \log_2\left(1 + \frac{|\mathbf{h}^\top \mathbf{w}|^2}{\sigma_s^2}\right) \\
    \st~&  \|\mathbf{w}\|^2  \le P_{\max}, \\
     & 0 \le \tilde{x}_{p,n} \le \tilde{x}_{\max}, p\in \mathcal{P},~ n \in \mathcal{N},
    \end{align}
\end{subequations}
where $\mathbf{h} = [\mathbf{h}_B,\mathbf{h}_G]$ denotes the joint channel from both the BS and the pinching antennas to the user. Furthermore, the joint beamforming vector is defined as $\mathbf{w} = [\mathbf{w}_B^\top, \mathbf{w}_G^\top]^\top$, where $\mathbf{w}_B $ and $ \mathbf{w}_G$ represents the beamforming vectors of the BS and the pinching antennas, respectively. 
Problem \eqref{eq:opt_cooperative_pinching} involves highly-coupled variables in both the objective function and the constraints. Additionally, location variables appear in the phase terms, which together render the problem non-convex and particularly challenging to solve.

Instead of solving problem \eqref{eq:opt_cooperative_pinching} directly, the authors of \cite{zhou2025joint} instead explored the characteristics of cooperative transmission to gain insights that can facilitate its solution. 
Adopting different levels of cooperation, the joint BS and the pinching-antenna transmission system can operate in three district modes: 1) standalone deployment (SD): the BS distributes its signal to multiple waveguide processing units (WPUs), each of which is deployed at a separate location. Each WPU independently feeds the signal to pinching antennas  along its associated waveguide, thereby enhancing the system coverage and communication performance; 2) semi-cooperative deployment (SCD): the BS transmits its signal to a centralized WPU located away from the BS. This  WPU then simultaneously distributes the signal to multiple waveguides, enabling coordinated transmission with moderate  complexity; 3) fully-cooperative deployment (FCD): the WPU is fully integrated in the BS, allowing the signal to be fed directly into the waveguides without additional intermediate interfaces. This tightly coupled configuration offers the highest level of integration and potentially the lowest transmission latency. 

Along with the three cooperative transmission modes in \cite{zhou2025joint}, power allocation in the SD and SCD schemes is performed in proportion to the number of RF chains at the BS and the waveguides for meeting the total transmit power budget across different transmission schemes. In the FCD scheme, the BS and pinching antennas perform joint signal processing and can flexibly allocate power among the $N_B + P$ RF chains based on the full channel state information (CSI) of both the channels from the BS to the users and the pinching antennas to the user. Following the pinching antenna placement scheme derived in \cite{xu2025rate}, the received SNR for different cooperative transmission schemes can be obtained in closed-form as shown in Table \ref{table:gathered_SNR}, where $L_G$ is the distance between the pinching antenna and the user, $\xi_P$ is the path loss coefficient, and $\tilde \gamma = \frac{P_{\max}}{\sigma_s^2}$ denotes the transmit SNR. The received SNR for joint transmission is influenced by multiple factors, and performance gains are not guaranteed under all parameter configurations. Hence, the choice of an appropriate cooperation scheme has to be tailored to the specific communication context.

\begin{table}
	\centering
	\caption{ Received SNR for different schemes \cite{zhou2025joint}}\label{table:gathered_SNR}
         \renewcommand{\arraystretch}{2.8}
	\begin{tabular}{cc}
            \hline
		\hline
		Schemes & Average Received SNR Expression \\
		\hline
		BS-only & ${\gamma _{BO}} = \frac{{\eta {N_B}}}{{L_B^{\xi_B }}}\tilde \gamma $  \\
		
		BS+PAS SD & ${\gamma _{SD}} = \left( {\frac{{\eta N_B^2}}{{L_B^{\xi_B }\left( {{N_B} + P} \right)}} + \frac{{\eta {N}P}}{{L_G^{\xi_P }\left( {{N_B} + P} \right)}}} \right)\tilde \gamma $ \\
		
		BS+PAS SCD & ${\gamma _{SCD}} = \left( {\frac{{\eta N_B^2}}{{L_B^{\xi_B }\left( {{N_B} + P} \right)}} + \frac{{\eta {N}{P^2}}}{{L_G^{\xi_P }\left( {{N_B} + P} \right)}}} \right)\tilde \gamma$ \\
		
		BS+PAS FCD & ${\gamma _{FCD}} = \left( {\frac{{\eta {N_B}}}{{L_B^{\xi_B }}} + \frac{{\eta {N}P}}{{L_G^{\xi_P }}}} \right)\tilde \gamma $ \\
		\hline
        \hline
	\end{tabular}
\end{table}

It is worth noting that deploying pinching antennas in existing cellular networks inherently involves multipoint cooperative transmission and support for multiple users, which leads to significantly more complex optimization problems. These include joint beamforming design, power allocation, scheduling, and user association, all of which collectively increase the problem’s dimensionality and computational complexity. Deployment strategies aimed at maximizing the performance gains of multi-user pinching-antenna systems have been studied in \cite{xu2025pinching,wang2025antenna,qian2025pinching}, and can be integrated with coordinated multipoint systems \cite{Hu2010joint,Andrea24coherent} to achieve further performance improvements. Therefore, effective optimization is critical for fully realizing the benefits of cooperative transmission in practical network deployments, particularly with regard to spectral efficiency, coverage enhancement, and scalability.

\subsection{Synergizing Pinching Antennas with Other Advanced Techniques}
As wireless networks continue to evolve, the integration of pinching-antenna systems with other advanced physical-layer techniques has emerged as a key strategy for addressing the diverse performance, security, and efficiency demands of next-generation applications. By harnessing the spatial flexibility and reconfigurability of pinching antennas, researchers have explored powerful synergies between this technology and state-of-the-art solutions in physical layer security, wireless power transfer, and intelligent wireless services. In this subsection, representative works are surveyed to illustrate the breadth and transformative potential of pinching-antenna systems when combined with cutting-edge wireless technologies.

\subsubsection{Pinching Antennas-Assisted Physical Layer Security}
By leveraging the unique capability of dynamically configuring antenna positions along dielectric waveguides, pinching-antenna systems have recently emerged as a promising paradigm for enhancing physical layer security in wireless networks. The spatial flexibility introduces new degrees of freedom for secure transmission, allowing transmitters to form strong LoS links to legitimate users while spatially distancing or nulling eavesdroppers. Recent studies have explored a range of secure transmission strategies for pinching-antenna systems \cite{sun2025physical,badarneh2025physical,papanikolaou2025secrecy,zhu2025pinching,wang2025pinching_pls,lu2025dual,jiang2025pinching}. For instance, the work \cite{sun2025physical} formulated joint baseband and pinching beamforming optimization problems to maximize the secrecy rate and proposed gradient-based and FP-BCD algorithms, showing that the pinching-antenna systems outperform conventional fixed-antenna systems as well as advanced linear precoding benchmarks in both single-user and multi-user wiretap scenarios. On the other hand, a rigorous performance analysis for single-user systems was conducted in \cite{badarneh2025physical}, where closed-form expressions for the average secrecy capacity and secrecy outage probability were derived. Besides, the results also demonstrated that placing pinching antennas close to legitimate users and increasing spatial separation from eavesdroppers significantly enhances secrecy metrics. Other works have further extended the study to artificial noise injection \cite{papanikolaou2025secrecy}, and secure multi-user MIMO scenarios \cite{zhu2025pinching}. In \cite{lu2025dual}, the authors investigated physical-layer security in dual-waveguide pinching-antenna systems, where multiple pinching antennas act as flexible channel reconfigurators to enhance secure sum rate and energy efficiency, and proposed a two-stage optimization algorithm that jointly designs antenna placement, beamforming, and artificial noise under parallel and orthogonal waveguide deployments.
In addition, the pinching-antenna-assisted covert communications have been studied in \cite{jiang2025pinching,he2025multicast}. In particular, the authors of \cite{jiang2025pinching} leveraged the spatial reconfigurability of pinching-antenna systems to boost covert communication capabilities, demonstrating that dynamic antenna positioning and beamforming can maximize covert rates and enhance transmission stealthiness in the presence of adversaries. While the authors of \cite{he2025multicast} studied the pinching-antenna-enabled covert communications in ISAC systems.
Overall, the existing research results show that pinching-antenna-enabled flexible antenna placement, joint secure beamforming, and waveguide-based LoS links together provide powerful tools for improving secrecy performance in future wireless systems, though challenges remain in real-time optimization, practical hardware design, and the integration of security with other advanced wireless functionalities.

\subsubsection{Pinching Antennas for Wireless Power Transfer}
Pinching-antenna systems have recently gained attention as a powerful tool for overcoming critical limitations in wireless power transfer (WPT) and energy harvesting networks. Traditional WPT systems often suffer from the ``double near-far'' problem, where users far away from the transmitter receive both less energy and have weaker communication channels. By enabling the flexible activation and placement of antennas along waveguides, pinching-antenna systems allow  transmitters to deliver energy much closer to distant users, significantly improving energy harvesting efficiency and uplink throughput. For example, the work in\cite{papanikolaou2025resolving} proposed a wireless powered pinching-antenna network, where distributed pinching antennas are jointly optimized for downlink energy transfer and uplink data collection. This architecture enhances user fairness and network performance compared to conventional fixed-antenna solutions. Building on these ideas, further works have investigated resource allocation, optimal antenna activation, and joint power optimization in WPT-enabled pinching-antenna systems, using methods such as game theory and particle swarm optimization to efficiently handle the associated non-convex problems \cite{li2025sum,peng2025computation}. The wireless powered mobile edge computing in pinching-antenna system was investigated in \cite{liu2025wireless}.

Moving beyond pure power transfer, pinching-antenna technology has also been leveraged to realize simultaneous wireless information and power transfer (SWIPT) in next-generation wireless networks \cite{li2025mimo,li2025pinching,jiang2025spatially,zhang2025performance}. These works jointly optimize power allocation and antenna positioning to balance the trade-off between data throughput and harvested energy, demonstrating substantial performance improvements over conventional fixed-antenna SWIPT architectures. Various algorithmic frameworks, such as particle swarm optimization and alternating optimization methods, have been employed to maximize the overall system efficiency.

\subsubsection{More Applications of Pinching Antennas}
Recent advances have also demonstrated that pinching-antenna systems hold significant promise for a wide range of emerging wireless applications beyond classical communications. For example, the authors of \cite{yang2025pinching2} introduced a novel pinching-antenna-assisted index modulation scheme, where both conventional modulation symbols and the indices of activated pinching-antenna positions are used for information delivery. This approach enables higher spectral efficiency without increasing hardware complexity, and substantially outperforms conventional modulation systems, highlighting the versatility and potential of pinching antennas in enabling advanced modulation and coding techniques. Pinching antenna-assisted symbiotic radio was studied in \cite{wang2025joint}, aiming to enhance both primary and secondary transmissions by leveraging the flexibility of pinching antennas. The authors of \cite{zhang2025pinching} investigated the pinching-antenna enabled indoor positioning.

In summary, the remarkable flexibility and spatial adaptability of pinching-antenna systems have opened up a diverse landscape of innovative applications. By enabling the dynamic configuration of antenna positions and radiation patterns, pinching antennas can address critical challenges in reliability, security, energy efficiency, and wireless intelligence that are increasingly crucial to next-generation networks. As research continues to explore new use cases and develop practical solutions, pinching-antenna technology is poised to play a pivotal role in shaping the evolution of future wireless systems, offering unprecedented opportunities for both foundational advancements and real-world deployment.

\section{Charting the Future: Open Questions and Research Directions} \label{sec: future work}
In this section, we outline several important future research directions that are essential for improving the performance and practicality of pinching-antenna systems.

\subsection{Pinching-Antenna System Design under Random LoS/NLoS Channels}
LoS blockage is a major challenge in high-frequency wireless systems, particularly in complex indoor environments, dense urban areas, and tunnel-like structures. Obstacles and user mobility can obstruct the direct propagation path between transmitters and receivers, leading to significant signal attenuation and degraded communication performance. These effects are especially pronounced in systems that rely on direct-path transmission. As pinching-antenna systems are envisioned for deployment in such environments, understanding, mitigating, and exploiting the impact of both LoS blockage and NLoS becomes essential for robust system design.

Recent research has begun to investigate how pinching-antenna systems can be adapted to handle LoS blockage. In particular, the authors of \cite{ding2025los} revealed a counterintuitive insight: the presence of LoS blockage can actually enhance the relative performance gain of pinching antennas over conventional ones. This benefit becomes especially prominent in multi-user settings, where LoS blockage naturally suppresses co-channel interference and amplifies the pinching antenna's spatial flexibility. The study in \cite{xu2025pinching_los} considered a multi-user pinching-antenna system under probabilistic LoS blockage and studied the average data rate maximization problem. Moreover, the authors of \cite{wang2025pinching_los} leveraged selective antenna activation to dynamically construct LoS links or rely on NLoS paths for interference mitigation. These results highlight the flexibility of pinching antennas in adapting to blocked environments, by enabling dynamic reconfiguration to steer around obstacles or exploit reflected paths—capabilities not easily achievable with fixed antenna arrays.

Initial studies on pinching-antenna system design under random LoS/NLoS channels in \cite{xu2025pinching_nlos} have already shown that pinching antennas can provide significant gains in such challenging propagation environments, see Section \ref{sec: los_nlos}. However, several open challenges merit further investigation, including scalable designs for multi-waveguide pinching-antenna topologies, robust optimization algorithms that are resilient to partial or uncertain blockage, and real-time reconfiguration schemes that can adapt to highly dynamic environments. Addressing these issues will be key to fully unlocking the potential of pinching-antenna systems in next-generation wireless networks operating under severe propagation constraints.

\subsection{Realization and Optimization of Efficient Pinching-Antenna Radiators} 
While the concept of pinching antennas offers tremendous deployment flexibility, the realization of efficient and controllable radiators remains a major open challenge. Most existing works rely on simple ``clothespin-style'' mechanical prototypes, which, though intuitive and easy to implement, are unlikely to yield optimal performance in terms of radiation efficiency, pattern control, and robustness.

There is an urgent need for systematic studies on the optimal shape, size, material, and placement of pinching-antenna radiators. Key questions include: How can radiator geometry and material be customized for maximal efficiency and minimal energy loss? What positioning strategies best enhance link quality and minimize mutual coupling between adjacent antennas? Is it possible to enable real-time or electronic reconfiguration of radiation patterns (beyond simple repositioning) for advanced beamforming, interference suppression, and multi-user support? Furthermore, practical realization will require methods to suppress undesired side effects, such as accidental leakage, shadowing, or inconsistencies due to manufacturing tolerances.

Addressing these questions may benefit from advances in metasurface design, additive manufacturing, and electronically tunable materials. Future work should also explore integration with active circuit elements, facilitating functions such as beam steering, or even reconfigurable array topologies. Ultimately, realizing efficient and controllable pinching antennas will be key to unlocking their full potential in next-generation wireless systems.

\subsection{Pinching-Antenna-Enabled Near-Field Communications}

Near-field communication is emerging as an important direction for future wireless networks, owing to its ability to exploit the unique properties of spherical wavefronts \cite{liu2023near,an2024near}. This paradigm enables highly focused energy transmission, enhanced spatial multiplexing, and the potential for distance-dependent beamforming—capabilities that are unattainable with conventional far-field (i.e., plane-wave) models \cite{zhang2022beam,ding2023resolution,han2023toward,ding2023noma}. However, realizing these benefits necessitates antenna arrays with sufficiently large apertures to ensure that users reside within the near-field region (i.e., within the Rayleigh distance), where the effects of spherical propagation become significant. Traditional antenna technologies often fall short in this regard due to constraints related to physical size, deployment flexibility, and cost.

Pinching-antenna systems present an attractive solution to these challenges. Their inherent ability to dynamically activate and precisely position antennas along long dielectric waveguides makes it possible to construct large-aperture arrays spanning several meters, tailored to specific user or application requirements. This flexibility and scalability enable the practical exploitation of near-field effects, overcoming the limitations of conventional approaches and facilitating more adaptive and efficient wireless environments.

Looking to the future, several key research challenges need to be addressed for pinching-antenna-enabled near-field communications. These include the development of accurate channel estimation methods that account for spherical-wave propagation, and the design of beamforming and user association algorithms that fully leverage near-field characteristics. Addressing these issues will be essential to fully realize the potential of pinching-antenna-enabled near-field communications in future ultra-dense, adaptive, and high-capacity wireless networks.

\subsection{Accurate and Realistic Electromagnetic Modeling of Pinching Antennas} 
Despite the growing interest in pinching-antenna systems, most analytical and system-level studies rely on highly idealized electromagnetic models. Often, models assume omnidirectional radiation, lossless coupling, and negligible impacts of antenna shape. In real-world implementations, however, pinching antennas are expected to exhibit significant deviations from such assumptions, including directionality, frequency-and-material-dependent loss, complex coupling, and manufacturing-induced irregularities.

Accurate and credible modeling is necessary not only to guide the design of efficient pinching antennas but also to enable credible system optimization and performance evaluation. Key research problems include: How do different radiator shapes, attachment mechanisms, and materials impact the actual radiation pattern and efficiency? What are the effects of mutual coupling, finite conductivity, and real waveguide geometry on signal propagation and system capacity? How can we realistically model losses in analytical expressions and simulations? What are the best methods (e.g., full-wave simulation, empirical calibration) to verify and refine such models?

Progress in this area will require the development of full-field electromagnetic simulation frameworks, experimental validation platforms, and parametric models that connect antenna geometry, material, and environment to real-world performance metrics. Moreover, hybrid modeling approaches that blend analytical tractability with empirical realism will be essential for bridging the gap between prototype-level understanding and optimized system deployment. Advancing electromagnetic modeling is thus a foundational priority for bringing pinching-antenna systems from concept to practice in future wireless networks.

\subsection{Environment-Aware Pinching-Antenna System Designs}

Environment-aware designs, which consider factors such as user data traffic patterns and obstacle patterns, are crucial for optimizing coverage, capacity, and reliability in dynamic and complex environments. This is particularly attractive for outdoor or highly dynamic scenarios where per-user tracking is difficult. Notably, environment-aware designs typically facilitate long-term network planning, as the environment generally does not change rapidly. Such designs allow networks to adapt to environmental conditions, ensuring robust performance even under challenging scenarios. However, traditional antenna systems are often limited by their fixed deployment, lacking the adaptability to respond to changing user distributions and evolving environmental factors.

Pinching-antenna systems are particularly well-suited for environment-aware designs due to their inherent flexibility. By adjusting the positions of antennas along pre-deployed waveguides, pinching antennas can dynamically respond to shifting user demands and changing environments. This adaptability makes them ideal for scenarios where conventional static antennas would perform poorly. Whether responding to hotspots in high-traffic areas or navigating obstacles in complex indoor environments, pinching-antenna systems offer a unique ability to optimize network performance through reconfigurable antenna placement over the long term.

Despite their potential, several challenges remain for realizing the full benefits of environment-aware pinching-antenna systems. Key challenges include developing efficient optimization algorithms for dynamic antenna placement, integrating both traffic-aware and geometry-aware models, and ensuring system robustness under fluctuating environmental conditions. Overcoming these challenges will be crucial for unlocking the full potential of pinching-antenna systems in real-world, long-term network deployments.

\subsection{Pinching-Antenna-Enabled Extremely Large-Scale Antenna Array}
Extremely large-scale antenna array (ELAA) systems are emerging as a key technology to meet the ambitious performance demands of next-generation wireless networks \cite{xu2025distributed,lu2024tutorial}. By dramatically increasing the number of antennas, ranging from hundreds to thousands, ELAA systems can substantially improve spatial resolution, spectral efficiency, link reliability, and energy efficiency. ELAA can be realized through two primary approaches: (i) single-BS deployment, where a BS aggregates a vast array of antennas into a centralized structure; and (ii) coordinated distributed deployment, where multiple spatially separated access points, each equipped with a subset of antennas, are jointly coordinated to function as a virtual super-array. While both forms offer significant gains, they also face critical implementation challenges, including high interconnection costs, computational complexity, and practical issues such as synchronization and backhaul limitations \cite{xu2025distributed,xu2023low,xu2024joint}.

Pinching antennas present a promising enabler for practical ELAA deployment. In a typical architecture, multiple waveguides are deployed and each waveguide hosts a large number of pinching antennas along its length. This allows a very large number of radiating points to be realized from a moderate number of feeder points and RF chains, enabling compact massive arrays at a single BS and low-cost expansion of distributed antenna networks. A recent study on large-scale multi-cell deployment of pinching antennas  \cite{zhu2025topological} further confirms the potential of such architectures, showing that appropriately designed waveguide-based deployments can reshape cell boundaries, improve worst-case cell performance, and enhance network-wide data rates. By leveraging the strong LoS links and reduced path loss inherent to pinching-antenna architectures, such ELAA systems can offer enhanced channel quality and spatial selectivity while significantly reducing hardware and cabling complexity. In both single-BS and distributed ELAA scenarios, multiple waveguides with reconfigurable pinching antennas support scalable multi-user services via spatial multiplexing across waveguides and conventional resource-domain schemes (e.g., OMA/NOMA).

Going forward, the integration of pinching antennas into ELAA architectures opens several exciting research directions. These include the design of distributed signal processing and beamforming algorithms tailored to flexible pinching-antenna arrays, the development of robust calibration and synchronization techniques in dynamically reconfigurable environments, and the joint optimization of pinching-antenna positioning with resource allocation and user association. Overall, pinching-antenna-enabled ELAA systems offer a promising pathway toward practical, scalable, and cost-effective ELAA deployments.

\subsection{Pinching Antenna-Assisted Wireless Edge Intelligence}
Edge AI is envisioned as a key enabler for next-generation wireless networks \cite{zhang2024generative4,wang2023toward,zhang2025toward,wang2023beyond,zhang2025towardAgenAI}. By bringing data processing and model training closer to end user devices, wireless edge intelligence supports a wide range of intelligent applications such as federated learning, autonomous driving, smart manufacturing, and real-time environment perception. However, the performance of distributed AI algorithms in wireless networks is often limited by imperfect and dynamically changing propagation conditions, such as fading and interference \cite{wang2021quantized,wang2025robust,wang2025communication,wu2024client}. These impairments can significantly degrade the convergence, reliability, and latency of wireless model training and inference.

Pinching-antenna systems offer a unique opportunity to address these challenges by flexibly reconfiguring the wireless propagation environment. By dynamically positioning pinching antennas to create strong LoS links and mitigate large-scale path loss, it becomes possible to provide stable, high-quality communication channels between distributed edge devices and the cloud. This, in turn, can enhance the robustness and efficiency of wireless AI model training, reduce transmission delays, and improve overall system intelligence. Notably, recent studies have demonstrated that pinching-antenna systems can dramatically enhance over-the-air computation  performance by physically shaping channel characteristics to minimize computation errors, thus enabling more accurate and efficient data aggregation for wireless edge intelligence \cite{lyu2025pinching}.

Looking ahead, several important research directions remain open in this area. One promising avenue is the joint design of pinching-antenna placement and wireless resource allocation to enhance learning performance at the edge. Additionally, developing adaptive activation strategies that enable real-time, environment-aware reconfiguration of pinching antennas will be crucial for maintaining robust and efficient edge intelligence in dynamic wireless environments. Further investigation is also needed into privacy, security, and energy efficiency issues that may arise when integrating pinching-antenna systems with distributed AI. Overall, the integration of pinching-antenna technology and AI-driven wireless networks holds significant promise for enabling intelligent, resilient, and high-performing next-generation communication systems.

\subsection{Robust Pinching-Antenna System Design}

The performance of pinching-antenna systems is intrinsically linked to accurate user position information and CSI. In practical deployments, however, the precise user locations and accurate CSI are often unavailable due to limitations in positioning technologies, user mobility, and measurement errors \cite{laoudias2018survey,niu2018received}. These uncertainties, encompassing both user position errors and CSI uncertainty, can lead to significant performance degradation, as antenna placement, beamforming strategies, and resource allocation become suboptimal when errors are present.

Despite its practical importance, the robust design of pinching-antenna systems under user position and CSI uncertainty remains largely unexplored in the current literature. Addressing this open challenge will require the development of robust optimization frameworks that explicitly incorporate both user position uncertainty and CSI errors into the system design. Potential approaches may include maximizing worst-case performance metrics (such as the minimum data rate or SNR), minimizing outage probability, or leveraging stochastic models to represent these uncertainties \cite{wang2014outage,cui2018outage}. These problems are inherently complex, often demanding new algorithmic strategies that jointly optimize antenna positioning, resource allocation, and beamforming, while maintaining computational efficiency for dynamic and large-scale networks.

\subsection{Real-Time Low-Complexity Control of Mechanical Antenna Placement}

With the inherent flexibility of pinching-antenna systems comes the significant challenge of efficiently controlling antenna positions in real time, especially as network scale and user dynamics become more complex \cite{acarnley2002stepping}. High-precision mechanical repositioning not only increases computational requirements but also incurs mechanical costs, which can be prohibitive in large-scale or rapidly changing deployments. To address these challenges, several promising research directions can be pursued. First, discretizing potential antenna positions to a finite set of precomputed grid points along the waveguide transforms the problem from a high-dimensional, fine-grained optimization into a tractable combinatorial selection task, greatly reducing computational overhead. Second, joint utility maximization frameworks can be developed to consider both traditional communication performance metrics (such as SNR or throughput) and the cost associated with antenna placement, including energy consumption and time delay, thus enabling smarter, cost-aware repositioning strategies. Third, event-driven repositioning strategies, in which antennas are repositioned only in response to specific triggers (such as sudden drops in channel quality or the detection of user mobility), can further minimize unnecessary adjustments and reduce system overhead \cite{anastasi2009energy,heemels2012introduction}. Collectively, these approaches hold great promise for enabling scalable, responsive, and practical real-time control in pinching-antenna deployments, particularly in dynamic and heterogeneous wireless environments.

\subsection{Practical Implementation Challenges and Scalability}

Beyond the algorithmic and architectural issues discussed above, the real-world deployment of generalized pinching-antenna systems raises several practical challenges that remain largely unexplored. First, fabrication tolerances of the guiding media and pinching elements (e.g., variations in dielectric profile, LCX slot geometry, connector quality, and mechanical alignment) can affect in-medium attenuation, coupling efficiency, and phase consistency along the waveguide or conductor, and thus need to be captured by more realistic models. Second, large-scale deployments with multiple waveguides and many pinching points will require efficient calibration of both amplitude and phase, potentially in near-field operation and dynamic repositioning of radiating spots. Finally, scalability becomes a key concern when the number of reconfigurable points grows large, since control signaling, channel estimation, and coordination overhead can easily become bottlenecks. Addressing these issues will require joint hardware–algorithm co-design, measurement-driven modeling, and the development of scalable calibration and control schemes tailored to generalized pinching-antenna architectures. 

Complementing wired-fed architectures, recent work on wireless-fed pinching-antenna systems has opened up new possibilities for large-scale and multi-cell deployments \cite{wijewardhana2025wireless}. By feeding the waveguides through relay antennas instead of dedicated wired links, wireless-fed pinching-antenna systems preserve the core advantages of pinching antennas (guided propagation plus flexible radiating spots) while alleviating the cost, fragility, and routing constraints of long end-to-end waveguide connections. This makes them particularly attractive for multi-cell scenarios, where multiple BSs or relay nodes can cooperatively illuminate large areas using networks of pinching-enabled waveguides. At the same time, such architectures introduce additional practical challenges, including the joint design of BS–relay–waveguide topologies, management of inter-cell interference, synchronization across wirelessly fed segments, and scalable control of many distributed pinching points. Systematically addressing these issues is an important step toward truly massive, network-wide deployments of generalized pinching-antenna systems.

\section{Conclusion} \label{sec: conclusion}
Pinching-antenna systems offer a flexible and innovative solution to many of the limitations faced by conventional antenna technologies in next-generation wireless networks. By enabling dynamic antenna placement and activation along dielectric waveguides, pinching-antenna systems provide enhanced LoS connectivity, spatial adaptability, and strengthened support for advanced functionalities such as integrated sensing and communications.
In this paper, we have developed a generalized pinching-antenna framework, where different media, such as dielectric waveguides and LCX, different antenna activation methods can be employed. Then, we have introduced the underlying physical mechanisms and distinctive channel characteristics of representative pinching-antenna system architectures, encompassing configurations with both single and multiple waveguides, and discussed state-of-the-art system design strategies. We also reviewed the integration of pinching antennas with emerging wireless technologies and highlighted a broad range of promising application scenarios.
Despite these significant advances, practical deployment of pinching-antenna systems continues to face several challenges. Notable issues include the need for robust designs that account for user position uncertainty, the development of real-time, low-complexity hardware and control solutions, and deeper integration with distributed intelligence and edge computing paradigms. Addressing these open problems will be essential for fully realizing the benefits of generalized pinching-antenna systems in future wireless environments.
In summary, while substantial progress has been achieved, continued research and innovation are imperative to unlock the full potential of generalized pinching-antenna systems and to ensure their successful adoption in the evolving landscape of wireless communications.


\smaller[1]


\end{document}